\documentclass[a4paper,12pt]{article}
\pdfoutput=1
\usepackage{jheppub}

\usepackage{verbatim}

\usepackage{natbib}

\usepackage{amsmath,amsfonts,amssymb,mathrsfs}
\usepackage{color}
\usepackage{bbm}
\usepackage[normalem]{ulem}
\usepackage{enumitem}
\usepackage{yfonts}
\usepackage[latin1]{inputenc}
\usepackage{graphicx}
\usepackage{xspace}
\usepackage[font=small,labelfont=bf]{caption}
\usepackage{subcaption}

\usepackage[colorlinks=true]{hyperref} 
\hypersetup{
    bookmarks=true,         % show bookmarks bar?
    unicode=false,          % non-Latin characters 
    pdftoolbar=true,        % show Acrobat
    pdfmenubar=true,        % show Acrobat 
    pdffitwindow=false,     % window fit to page when opened
    pdfstartview={FitH},    % fits the width of the page to the window
    pdftitle={p-adic CFT is a Holographic Tensor Network},    % title
    pdfauthor={Ling-Yan Hung, Wei Li, and Charles M. Melby-Thompson},     % author
    pdfnewwindow=true,      % links in new window
    colorlinks=true,        % false: boxed links; true: colored links
    linkcolor=blue,         % color of internal links (change box color with linkbordercolor)
    citecolor=red,          % color of links to bibliography
    filecolor=magenta,      % color of file links
    urlcolor=cyan           % color of external links
}

\newcommand{\includeCroppedPdf}[2][]{%
    \IfFileExists{./#2-crop.pdf}{}{%
        \immediate\write18{pdfcrop #2 #2-crop.pdf}}%
    \includegraphics[#1]{#2-crop.pdf}}

\newcommand{\be}{ \begin{equation}}
\newcommand{\ee}{\end{equation}}
\newcommand{\ba}	{\begin{align}}
\newcommand{\ea}	{\end{align}}

\begin{document}
%%%%%%%%%%%%%%%%%%%%%%%%%%%%%%%%%%%%%%%%%%%%%%%%%%%%%%%%%%%%%%%%%%
\title{\bf Wilson line networks in $p$-adic AdS/CFT}
\author{Ling-Yan Hung$^{a,b,c}$,  Wei Li$^{d}$ and Charles M. Melby-Thompson$^{e}$}

\affiliation{$^a$ Department of Physics and Center for Field Theory and Particle Physics,\\
\hspace*{0.3cm}Fudan University, Handan Road, 200433 Shanghai, P.\ R.\ China}

\affiliation{$^b$ State Key Laboratory of Surface Physics and Department of Physics,\\ 
\hspace*{0.3cm}Fudan University, 220 Handan Road, 200433 Shanghai, P.\ R.\ China}
        
\affiliation{$^c$ Collaborative Innovation Center of Advanced  Microstructures,\\
\hspace*{0.3cm}Nanjing University, Nanjing, 210093, P. R. China.}
        
\affiliation{$^d$ Institute of Theoretical Physics,\\
\hspace*{0.3cm}Chinese Academy of Sciences, 100190 Beijing, P.\ R.\ China}

\affiliation{$^e$ Institut f\"{u}r Theoretische Physik und Astrophysik,\\
\hspace*{0.3cm}Julius-Maximilians-Universit\"{a}t W\"{u}rzburg, Am Hubland, 97074 W\"{u}rzburg, Germany}

\emailAdd{elektron.janethung@gmail.com, 
weili@itp.ac.cn, charlesmelby@gmail.com}\abstract{
The $p$-adic AdS/CFT is a holographic duality based on the $p$-adic number field $\mathbb{Q}_p$. 
For a $p$-adic CFT living on $\mathbb{Q}_p$ and with complex-valued fields, 
 the bulk theory is defined on the Bruhat-Tits tree, which can be viewed as the bulk dual of $\mathbb{Q}_p$.
We propose that bulk theory can be formulated as a lattice gauge theory of PGL$(2,\mathbb{Q}_p)$ on the Bruhat-Tits tree, and show that the Wilson line networks in this lattice gauge theory can reproduce all the correlation functions of the boundary $p$-adic CFT. 
}

\maketitle
\flushbottom

%%%%%%%%%%%%%%%%%%%%%
%\newpage

\tableofcontents

\section{Introduction}

The $p$-adic AdS/CFT proposed in \cite{Heydeman:2016ldy, Gubser:2016guj} is a holographic duality between conformal field theories based on the $p$-adic number field $\mathbb{Q}_p$ and bulk theories living on the Bruhat-Tits tree \cite{BruhatTits} --- a $(p+1)$-valent tree that can be viewed as the bulk dual of $\mathbb{Q}_p$ \cite{Zabrodin:1988ep}.
It shares many important features of ordinary AdS/CFT based on the real field $\mathbb{R}$. 
The field  $\mathbb{Q}_p$ results from completing the rational field $\mathbb{Q}$ using the $p$-adic norm rather than the Euclidean norm. 
In fact, it is the only field other than $\mathbb{R}$ that can be obtained by completing $\mathbb{Q}$ while subject to the four axioms of Euclidean norm \cite{Ostrowski}.
The generalization of  AdS/CFT from $\mathbb{R}$ to $\mathbb{Q}_p$ suggests that the holography is more universal than in spacetime based on $\mathbb{R}$.
There has been many recent developments in the last two years, see. \cite{Gubser:2016htz, Gubser:2017vgc, Gubser:2017tsi, Dutta:2017bja, Gubser:2017qed, Qu:2018ned, Stoica:2018zmi, Jepsen:2018dqp,   Gubser:2018cha, Gubser:2018ath}.

Tensor network originally started as an ansatz for solving $N$-body wavefunctions (see e.g.\ review \cite{Orus:2013kga} and references therein) and recently has been used to realize discrete versions of holographic duality \cite{Pastawski:2015qua, Hayden:2016cfa}, where the tensor network lives on a discretized bulk spacetime (hence the bulk isometry is broken to a discrete subgroup of the conformal group).  
Many important features of the AdS/CFT dictionary, most notably Ryu-Takayanagi formula and the structure of the entanglement wedge, are naturally captured by a suitable tensor network (at least qualitatively) \cite{Swingle:2009bg}. 
This suggests that tensor network might help uncover the mechanism of AdS/CFT correspondence.

In \cite{Bhattacharyya:2017aly},  we proposed that one can use a tree-type tensor network living on the Bruhat-Tits tree  to provide a concrete realization of $p$-adic AdS/CFT. 
In this realization, the dictionary between the boundary and bulk sides of the $p$-adic AdS/CFT is derived from the tensor network instead of being treated as a conjecture.
In particular, we have derived the bulk reconstruction formula  and shown how to recover boundary correlation functions from the tensor network.\footnote{For the corresponding results for AdS/CFT based on $\mathbb{R}$, see e.g.\ \cite{Hamilton:2005ju, Hamilton:2006az, Hamilton:2006fh} for the bulk reconstruction formula and see \cite{Gubser:1998bc, Witten:1998qj} for Witten diagram computations of boundary correlation functions.}
In a more recent work \cite{us2}, we reproduced explicitly the complete set of correlation functions of any $p$-adic CFT with given spectra and structure constants. 

In the $p$-adic AdS/CFT discussed so far, boundary correlation functions are reproduced by bulk Witten diagrams. 
In AdS$_3$/CFT$_2$ based on $\mathbb{R}$, the bulk Einstein gravity can be reformulated as an SL$(2,\mathbb{C})$ Chern-Simons theory \cite{Witten:1988hc, Witten:2007kt} and accordingly the boundary correlation functions can also be reproduced by Wilson line networks of the Chern-Simons theory \cite{Bhatta:2016hpz,Besken:2016ooo,Fitzpatrick:2016mtp}.
For AdS$_3$/CFT$_2$, the Wilson line network computation of boundary correlation functions is simpler and conceptually more elegant, e.g.\ there is no need to integrate the bulk point of the Witten diagram and the $1/c$ expansion is more transparent. 
Therefore it is natural to ask whether for $p$-adic AdS/CFT, the  bulk theory could also have an alternative description in terms of a Chern-Simons like theory.\footnote{Note that for $p$-adic AdS/CFT, this alternative formulation exists for all dimensions, in contrast to AdS/CFT based on $\mathbb{R}$.
Different from AdS$_{d+1}$/CFT$_d$ for $\mathbb{R}$, the spacetime dimension in $p$-adic AdS/CFT is realized as between CFT living in $\mathbb{Q}_{p^d}$ (instead of $(\mathbb{Q}_p)^{\otimes d}$) and the bulk theory on the corresponding extension of Bruhat-Tits tree. 
Therefore the bulk isometry is always the PGL$(2)$ group, with the only $d$-dependence being on the field $\mathbb{Q}_{p^d}$ \cite{Heydeman:2016ldy, Gubser:2016guj}.}

The fact that there should exist an alternative Chern-Simons like bulk theory for the $p$-adic AdS/CFT is also evident from the tensor network realization of the $p$-adic AdS/CFT.
In \cite{Bhattacharyya:2017aly, us2}, each tensor located at a vertex is proportional to the structure constants of the boundary CFT, suggesting a bulk gauge symmetry on the Bruhat-Tits tree that are directly furnished by the tensors. 
In addition, the Witten diagrams used to reproduce boundary correlation function actually do not involve a sum over positions of the interaction vertices.
These strongly suggest a connection with Wilson line network of some PGL$(2,\mathbb{Q}_p)$ gauge theory living on the Bruhat-Tits tree.\footnote{We thank Jieqiang Wu for suggesting to us the similarity between tensor networks and Wilson line networks.}
We will argue that the tensor network can be naturally interpreted as a Wilson line network of some PGL$(2,\mathbb{Q}_p)$ gauge theory. 
This is the analogue of the Wilson line networks in the SL$(2,\mathbb{C})$ Chern-Simons formulation of AdS$_3$ gravity, see e.g.\ \cite{Ammon:2013hba, Besken:2016ooo, Bhatta:2016hpz, Fitzpatrick:2016mtp, Castro:2018srf, Besken:2018zro, Kraus:2018zrn}.

We do not know of any Chern-Simons theory defined on the Bruhat-Tits tree.
To proceed, we draw on parallels with topological lattice gauge theories and also hints from the $\mathfrak{sl}(2)$ Chern-Simons formulation of Einstein gravity in AdS$_3$ to prescribe the basic ingredients of the PGL$(2,\mathbb{Q}_p)$ lattice gauge theory living on the Bruhat-Tits tree.
We will show that to produce the boundary correlation function of the $p$-adic AdS/CFT with complex-valued fields,\footnote{The $p$-adic AdS/CFT studied so far all have complex-valued fields, which have no descendants. 
On the other hand, $p$-adic CFT with $p$-adic valued fields allows descendants and is correspondingly richer. 
It would be very interesting to formulate $p$-adic AdS/CFT for  $p$-adic CFT with $p$-adic valued fields, and to generalize the results in this paper to that context.}
we actually only need a particular pure gauge configuration (analogous to the solution in the $\mathfrak{sl}(2,\mathbb{C})$ Chern-Simons theory that corresponds to the pure AdS$_3$ background).
We will then show how the Wilson line network with this PGL$(2,\mathbb{Q}_p)$ gauge connection reproduces all correlation functions of the boundary $p$-adic AdS/CFT.
We leave a complete formulation of the action and the dynamics of this theory to future work.

The paper is organized as follows. 
After a short review of $p$-adic AdS/CFT in Section 2, we explain basic ingredients on the Chern-Simons like lattice gauge theory living on the Bruhat-Tits tree and define its Wilson line networks. 
In Section 4, we show how these Wilson line networks reproduce correlation functions of the boundary $p$-adic CFT.
In Section 5, we give a tensor network realization of these Wilson line networks.
Finally Section 6 contains a summary and discussion on open problems, and Appendix A contains the derivation of a few properties of $p$-adic shadow operators.

\section{Review of $p$-adic AdS/CFT}

In this section we review some basic feature of $p$-adic number field and $p$-adic AdS/CFT that we will need later. For textbook on $p$-adic number and $p$-adic analysis, see e.g.\ \cite{Koblitz, Gouvea}. For works on $p$-adic number in string theory see e.g.\ \cite{Freund:1987kt, Freund:1987ck, Brekke:1988dg, Dragovich:2007wb}.

\subsection{$p$-adic number and $p$-adic analysis}

Start from the field of rational numbers $\mathbb{Q}$, one can impose the Euclidean norm $|x|$, which satisfies the following axioms
\begin{equation}\label{fouraxioms}
\begin{aligned}
&(1):\,\, |x|\geq 0 \qquad (2):\,\, |x|=0 \leftrightarrow x=0\qquad (3):\,\,  |x\, y|=|x|\, |y| 
\\
&(4):\,\, |x+y| \leq |x|+ |y|  \qquad 
\textrm{(triangle inequality)}
\end{aligned}
\end{equation}
Using the Euclidean norm to complete $\mathbb{Q}$, one obtains the field of real numbers $\mathbb{R}$.

It is possible to define another norm that satisfies all four axioms (\ref{fouraxioms}). Given a prime number $p$, a rational number $x\in \mathbb{Q}$ can always be written into
\begin{equation}
x = p^e \frac{u}{d} \qquad\textrm{with} \qquad e, u, d \in \mathbb{Z}
\end{equation}
and $u, d\in \mathbb{Z}$ and are not divisible by $p$. 
(Note that $e$ defined this way is unique.)
The $p$-adic norm of $x$ is defined as
\begin{equation}\label{padicnormdefrational}
|x|_p =\frac{1}{p^e}
\end{equation}
One can check that the norm defined by (\ref{padicnormdefrational}) obey all four axioms (\ref{fouraxioms}). In fact, it satisfies an even stronger version of the fourth axiom:
\begin{equation}\label{STI}
|x+y|_p \leq \textrm{max}(|x|_p,|y|_p) \qquad \textrm{(strong triangle inequality)}
\end{equation}

One can now complete the rational field $\mathbb{Q}$ using the $p$-adic norm $|x|_p$. 
The resulting field is called $p$-adic number field $\mathbb{Q}_p$. 
Given a prime number $p$, the field $\mathbb{Q}_p$ consists of all expansions of the form:
\begin{equation}\label{QpDef}
\mathbb{Q}_p \equiv \{x=\sum^{\infty}_{n=-N} a_n \, p^n \,\, |\,\,a_n \in \mathbb{F}_p\}.
\end{equation}
The definition of the $p$-adic norm (\ref{padicnormdefrational}) ensures that  the formal series (\ref{QpDef}) converges.
Finally, we emphasize that the Euclidean norm and the $p$-adic norm (for each prime $p$) are the only two types of norms that obey the four axioms \cite{Ostrowski}. Namely, the rational field $\mathbb{Q}$ has only two types of completions: $\mathbb{R}$ and $\mathbb{Q}_p$ (for each prime $p$).

\subsection{$p$-adic CFT}
\label{subsec: padicCFT}
The fields of $p$-adic CFT can be chosen to be either complex-valued or $p$-adic valued. 
We will focus on the former case, and only comment on the latter in the end. For more details, see \cite{Melzer:1988he}.

The global symmetry of the $p$-adic CFT is PGL$(2,\mathbb{Q}_p)$.\footnote{It is $\textrm{PGL}(2,\mathbb{Q}_p)$ instead of SL$(2,\mathbb{Q}_p)$ because the determinant $\sqrt{\det{g}}$ is not always in the field of $\mathbb{Q}_p$ (even though $g\in \textrm{GL}(2,\mathbb{Q}_p)$) --- therefore one cannot always use the projective equivalence to bring the determinant to one.}
In $p$-adic CFT with complex-valued fields, a primary field of weight $\Delta$ is defined to transform under the $p$-adic M\"obius transformation as \cite{Melzer:1988he}:
\begin{equation}\label{mobius}
\mathcal{O}_{\Delta}(x)\longrightarrow \tilde{\mathcal{O}}_{\Delta}(\frac{ax+b}{cx+d})= \left|\frac{ad-bc}{(cx+d)^2}\right|_p^{- \Delta} \mathcal{O}(x) \qquad \textrm{with} \quad \begin{pmatrix}
a& b\\
c& d
\end{pmatrix}\in \textrm{PGL}(2,\mathbb{Q}_p)
\end{equation}

Moreover, we will focus on ``locally constant" fields.
In $p$-adic analysis for complex-valued fields, the analogue to the ``smooth" function in real analysis is the ``locally constant" function. 
For a locally-constant function $\phi(x)$, there exists a slicing of the $p$-adic field into closed-open subsets  $\mathbb{Q}_p= \bigsqcup_{i} S_i$, such that $\phi(x)$ is constant within each $S_i$. It follows that a locally-constant function has zero-derivative:
\begin{equation}
\phi(x) \textrm{ is locally constant} \qquad \Longrightarrow \qquad \frac{\partial}{\partial x} \phi(x)=0
\end{equation}
This means that the complex-valued fields in $p$-adic CFT has no descendant --- all fields are (global) conformal primaries  \cite{Melzer:1988he}. 

Since there is no descendant, the OPE of two primaries is simply:
\begin{equation}\label{OPE}
 \mathcal{O}_1(x_1) \mathcal{O}_2(x_2)\sim\sum_{\ell} C_{12}^{\ell} \mathcal{O}_{{\ell}}(x_2)
\end{equation}
where the sum is over all primaries $\mathcal{O}_{\ell}$ that appear in the $\mathcal{O}_1 \mathcal{O}_2$ OPE.
This is in contrast to the meromorphic CFT: in $\mathcal{O}_{\ell}$'s channel of the $\mathcal{O}_1\mathcal{O}_2$ OPE, all descendants appear.

\subsubsection{Two and three-point functions}

As in the real case, the global conformal symmetry fixes the form of the $n$-point functions. The two-point functions are fixed to be
\begin{equation}
\langle \mathcal{O}_i(x_1) \mathcal{O}_j(x_2)\rangle=\frac{\delta_{ij}}{|x_{12}|^{2 \Delta_i}_{p}}
\end{equation}
where $x_{ij}\equiv x_i -x_j$ and we have ortho-normalized all fields accordingly. 
The three-point functions are fixed to be \cite{Melzer:1988he}
\begin{equation}\label{3ptMobius}
\begin{aligned}
\langle \mathcal{O}_1(x_1) \mathcal{O}_2(x_2)\mathcal{O}_3(x_3)\rangle=\frac{C_{123} }{ |x_{12}|^{\Delta_1+\Delta_2-\Delta_3}_{p}  |x_{23}|_p^{\Delta_2+\Delta_3-\Delta_1}  |x_{31}|^{\Delta_3+\Delta_1-\Delta_2}_p}
\end{aligned}
\end{equation}

For later discussion, it is instructive to check that this three-point function result (\ref{3ptMobius}) is consistent with the OPE expansion (\ref{OPE}). 
WLOG, we assume that $x_1$ and $x_2$ are closer to each other than to $x_3$:
\begin{equation}\label{3ptConfig}
\textrm{WLOG}:\qquad |x_{12}|_p \leq |x_{13}|_p \qquad \textrm{and} \qquad |x_{12}|_p \leq |x_{23}|_p
\end{equation}
therefore we should first compute the OPE between $\mathcal{O}_1$ and $\mathcal{O}_2$, which gives the three-point function
\begin{equation}\label{3ptOPE}
\begin{aligned}
\langle \mathcal{O}_1(x_1) \mathcal{O}_2(x_2)\mathcal{O}_3(x_3)\rangle_{\textrm{OPE}}& =\frac{C_{123}}{|x_{12}|^{\Delta_1+\Delta_2-\Delta_{3}}_p |x_{23}|^{2\Delta_3}}\\
\end{aligned}
\end{equation}
which at the first sight seems different from the result (\ref{3ptMobius}) dictated by the $p$-adic M\"obius symmetry.

However, the ``strong triangle inequality" (\ref{STI}) implies that every triangle in $\mathbb{Q}_p$ is an isosceles triangle whose leg is longer than its base:\footnote{This is a property that will be used extensively in this paper and  is responsible for many somewhat counter-intuitive features of $p$-adic CFT. 
} 
namely, the assumption (\ref{3ptConfig}) implies
\begin{equation}\label{isosceles}
\textrm{WLOG:} \qquad \qquad \left|x_{12}\right|_p \leq |x_{23}|_p= |x_{13}|_p 
\end{equation}
Therefore, the ratio between the three-point function that follows from the OPE (\ref{OPE}) and the one from the $p$-adic M\"obius symmetry (\ref{mobius}) is 
\begin{equation}
\frac{\langle \mathcal{O}_1(x_1) \mathcal{O}_2(x_2)\mathcal{O}_3(x_3)\rangle_{\textrm{OPE}} }{\langle \mathcal{O}_1(x_1) \mathcal{O}_2(x_2)\mathcal{O}_3(x_3)\rangle_{\textrm{M\"obius}} 
}=\left|\frac{x_{13}}{x_{23}}\right|^{\Delta_3+\Delta_1-\Delta_2}_p =1\end{equation}
From this exercise, one can also see that starting from three-point, the correlation function has a simpler form (i.e.\ (\ref{3ptOPE})) than its $\mathbb{R}$ cousin.
This feature is more prominent in the four-point and higher-point functions.

\subsubsection{Four-point function and global conformal block}

The four-point functions is constrained by the M\"obius symmetry to take the form
\begin{equation}\label{4ptpadic}
\begin{aligned}
&\langle \mathcal{O}_1(x_1) \mathcal{O}_2(x_2)\mathcal{O}_3(x_3)\mathcal{O}_4(x_4) \rangle\\
& \qquad \qquad =\frac{|\eta|_p^{\Delta_3+\Delta_4-\Delta/3} \, |1-\eta|_p^{\Delta_2+\Delta_3-\Delta/3}
}{\prod^{4}_{i<j} |x_{ij}|^{\Delta_i+\Delta_j-\Delta/3}_{p}}\langle \mathcal{O}_1(0) \mathcal{O}_2(\eta)\mathcal{O}_3(1)\mathcal{O}_4(\infty) \rangle \end{aligned}
\end{equation}
where $\Delta\equiv \sum^{4}_{i=1}\Delta_i$, the cross-ratio $\eta\equiv \frac{x_{12} \cdot x_{34}}{x_{13}\cdot x_{24}}$, and the conformal block \cite{Melzer:1988he}:
\begin{equation}\label{CB}
\begin{aligned}
&\langle \mathcal{O}_1(0) \mathcal{O}_2(\eta)\mathcal{O}_3(1)\mathcal{O}_4(\infty) \rangle \\
=&
\begin{cases}
&\sum_{\ell} \frac{C^{\ell}_{12} C^{\ell}_{34} }{|\eta|^{\Delta_1+\Delta_2-\Delta_{\ell}}_p}\qquad \qquad\quad \,\, |\eta|\leq |1-\eta|=1 \qquad \textrm{(s-channel)} \\
&\sum_{\ell} \frac{C^{\ell}_{14} C^{\ell}_{23} }{|1-\eta|^{\Delta_2+\Delta_3-\Delta_{\ell}}_p}\qquad\qquad \,\,\, |1-\eta|\leq |\eta|=1 \qquad \textrm{(t-channel)}\\
&\sum_{\ell} \frac{C^{\ell}_{13} C^{\ell}_{24} }{|\eta|^{-\Delta_1-\Delta_3+\Delta_{\ell}}_p}\qquad \qquad \quad1\leq |\eta| =|1-\eta|\qquad \textrm{(u-channel)}
\end{cases}
\end{aligned}
\end{equation}
where we have used the isosceles property.
The conformal block (\ref{CB}) is much simpler than its meromorphic counterpart. 
As a result, the constraints from the crossing symmetry are much weaker: 
\begin{equation}
\sum_{\ell}C_{ij\ell}C_{\ell km}=\sum_{\ell}C_{ik\ell}C_{\ell jm}=\sum_{\ell}C_{im\ell}C_{\ell kj}
\end{equation}
It is merely the associativity of the three-point coefficients  \cite{Melzer:1988he}!

\subsection{$p$-adic AdS/CFT}

In AdS/CFT, the boundary conformal symmetry should be the same as the isometry of the bulk AdS.\footnote{Except for AdS$_3$/CFT$_2$ where the bulk isometry $\mathfrak{sl}(2,\mathbb{R})$$\times$$ \mathfrak{sl}(2,\mathbb{R})$ is enhanced to $\textrm{Virasoro}$$\times$$ \textrm{Virasoro}$.}
For $p$-adic AdS/CFT, the bulk spacetime is given by the Bruhat-Tits tree. We will first review its definition and in particular its isometry group PGL$(2,\mathbb{Q}_p)$, which is the conformal symmetry of the boundary $p$-adic CFT.
Then we will briefly review essential features of the $p$-adic AdS/CFT.

\subsubsection{Bruhat-Tits Tree}

The Bruhat-Tits tree is the analogue of the upper half plane for the $p$-adic field $\mathbb{Q}_p$.
Recall that for the real field $\mathbb{R}$, the upper half plane is defined as $\mathbb{H}\equiv \mathrm{SL}(2,\mathbb{R})/\mathrm{SO}(2,\mathbb{R})$, where $\mathrm{SL}(2,\mathbb{R})$ is the isometry group of $\mathbb{H}$ and $\mathrm{SO}(2,\mathbb{R})$ its maximal compact subgroup. 
The real field $\mathbb{R}$ is the boundary of $\mathbb{H}$. 
The analogue of this $\mathbb{H}$ for $\mathbb{Q}_p$ is then simply 
\begin{equation}\label{Hcoset}
\mathbb{H}_p \equiv \frac{\textrm{PGL}(2,\mathbb{Q}_p)}{\textrm{PGL}(2,\mathbb{Z}_p)}.
\end{equation}
where $\textrm{PGL}(2,\mathbb{Q}_p)$ is the isometry group of this analogous upper half plane and $\textrm{PGL}(2,\mathbb{Z}_p)$ is the maximal compact subgroup of $\textrm{PGL}(2,\mathbb{Q}_p)$, where $\mathbb{Z}_p$ is the ring of $p$-adic integers: $\mathbb{Z}_p\equiv \{x\in\mathbb{Q}_p\,|\, |x|_p \leq 1\}$.

Since $\textrm{PGL}(2,\mathbb{Z}_p)$ is a ``clopen" (both close and open) subgroup of $\textrm{PGL}(2,\mathbb{Q}_p)$, $\mathbb{H}_p$ is discrete. 
In fact, it has the topology of a $(p+1)$ valent tree, whose boundary is the continuous field  $\mathbb{Q}_p$.\footnote{A lattice with continuous boundary is called Bethe lattice, and has many intriguing features.}
One can construct the tree $\mathbb{H}_p$ from its definition, in terms of equivalence classes of integer (i.e.\ $\mathbb{Z}_p$) lattices in $\mathbb{Q}_p\otimes \mathbb{Q}_p$ \cite{BruhatTits,Zabrodin:1988ep}.

A PGL$(2,\mathbb{Q}_p)$ transformation acts on a vector in $\mathbb{Q}_p\otimes \mathbb{Q}_p $ by
\begin{equation}
\vec{f}\longrightarrow   \vec{f}' =\gamma \cdot \vec{f}                 \quad \textrm{with}\quad \gamma
                 \in \textrm{PGL}(2, \mathbb{Q}_p ).
\end{equation}
Now consider the space of lattices $\langle\vec{f},\vec{g}\rangle$ in $\mathbb{Q}_p\otimes \mathbb{Q}_p$. 
The PGL$(2,\mathbb{Q}_p)$ acts transitively on this space. 
On the other hand, the stabilizer of a lattice $\langle\vec{f},\vec{g}\rangle$  is PGL$(2,\mathbb{Z}_p)$. 
Therefore, the Bruhat-Tits tree as defined by the coset (\ref{Hcoset}) is identical to the set of equivalence classes $\langle\langle\vec{f},\vec{g}\rangle\rangle$, where the equivalence is defined as 
\begin{equation}\label{isotropy}
        \begin{aligned}
\langle \vec{f},\vec{g} \rangle \sim \langle \vec{f}',\vec{g}' \rangle \qquad \textrm{iff} \quad          ( \vec{f}',\vec{g}' )
                =  ( \gamma \cdot \vec{f}, \gamma \cdot \vec{g}  ) 
                 \quad \textrm{with}\quad \gamma
                 \in \textrm{PGL}(2, \mathbb{Z}_p ).
        \end{aligned}
\end{equation}

\begin{figure}[h!]
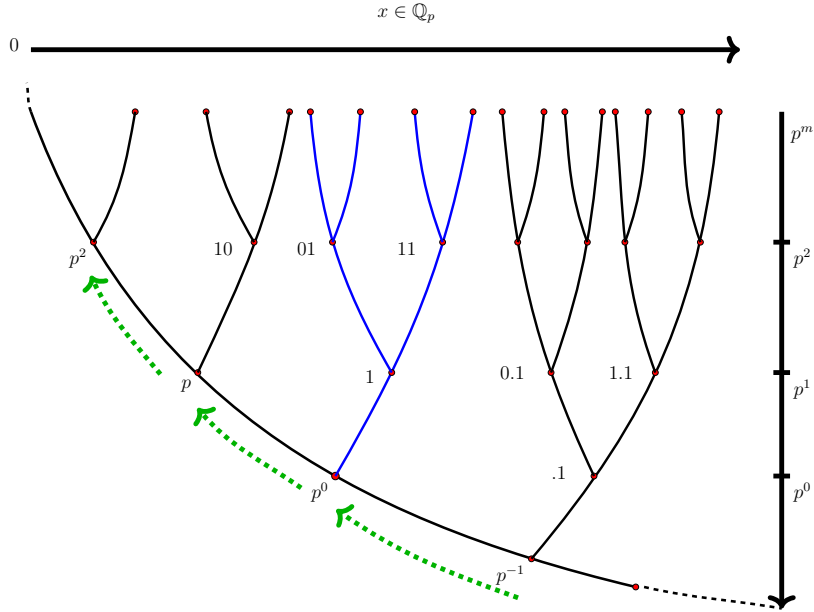

        \centering
        \includeCroppedPdf[width=0.7\textwidth]{"tess20"}
        \caption{Bruhat-Tits tree for $p=2$. The main branch is colored blue.}
        \label{fig:p2bttree}
\end{figure}

Now we explain the coordinate system on $\mathbb{H}_p$ that resembles the one on  the upper half plane $\mathbb{H}$.
The origin is chosen to be
\begin{equation}\label{treeorigin}
o\equiv \langle \langle \vec{f}_0, \vec{g}_0\rangle\rangle \qquad \textrm{with}\qquad         
\vec{f}_0 \equiv \left(\begin{matrix}1\\ 0 \end{matrix}\right) \qquad   
\vec{g}_0 \equiv \left(\begin{matrix}0\\ 1 \end{matrix}\right)   . 
\end{equation}
The $(p+1)$ neighbors of $o$ are
\begin{equation}\label{nnO}
\langle \langle\left(\begin{matrix}1\\ 0 \end{matrix}\right), \left(\begin{matrix}0\\ p \end{matrix}\right)\rangle\rangle \qquad \qquad \langle \langle\left(\begin{matrix}p\\ 0 \end{matrix}\right), \left(\begin{matrix}n\\ 1 \end{matrix}\right)\rangle\rangle  \qquad \textrm{with}\quad n=0,1,\dots, p-1.
\end{equation}  
which can be generated by the $(p+1)$ matrices in the Hecke operator
\begin{equation}\label{Heckes}
\Gamma_{-1} \equiv \left(\begin{matrix}1 &0\\ 0&p \end{matrix}\right) \qquad \textrm{and} \qquad\Gamma_{n}\equiv \left(\begin{matrix}p &n\\ 0&1 \end{matrix}\right)\qquad n=0,1,\dots,p-1
\end{equation}
on the vertex $\langle \langle \vec{f}_0, \vec{g}_0\rangle\rangle$. 
Then we use the projective equivalence to fix the second vector $\vec{g}$ to $\vec{g}_0$, i.e.\ rescale the first vertex in (\ref{nnO}) into $\langle \langle\left(\begin{matrix}\frac{1}{p}\\ 0 \end{matrix}\right), \left(\begin{matrix}0\\ 1\end{matrix}\right)\rangle\rangle$.
Applying the $(p+1)$ operators in (\ref{Heckes}) on the origin (\ref{treeorigin}) of the tree iteratively then generates the entire tree, with all vertices having the form
\begin{equation}\label{Hpcoord}
        \langle \langle \left(\begin{matrix}p^m\\ 0 \end{matrix}\right), \left(\begin{matrix}x^{(m)}\\ 1\end{matrix}\right)\rangle\rangle
  \qquad \qquad x^{(m)}=\sum^{m-1}_{ n=-N} a_n p^n \qquad a_n \in\mathbb{F}_p ,  \end{equation}
where again we have used the projective equivalence to set $g_2=1$. 
The Bruhat-Tits tree for $p=2$ is shown in Figure \ref{fig:p2bttree}, where we have labeled some vertices using 
coordinate system (\ref{Hpcoord}). 
Finally, since $x^{(m)}$ truncates at $p^{m}$,  we can regard $p^m$ as imposing the accuracy level of the $p$-adic number $x^{(m)}$, i.e.\ the vertex (\ref{Hpcoord})
represents the equivalence class $x^{(m)}+p^m \mathbb{Z}_p$.

\subsubsection{$p$-adic AdS/CFT}

We have just seen how the Bruhat-Tits tree furnishes the  isometry PGL$(2,\mathbb{Q}_p)$, which is the M\"obius symmetry PGL$(2,\mathbb{Q}_p)$ of the $p$-adic CFT. 
Other aspects of $p$-adic AdS/CFT, such as the correspondence between bulk fields living on the Bruhat-Tits tree and the boundary operators, can be found in  \cite{Heydeman:2016ldy, Gubser:2016guj}. 
In particular, after defining actions of bulk fields living on the Bruhat-Tits tree,  one can then recover boundary correlation functions by computing Witten diagrams, in a way analogous to the usual AdS/CFT based on $\mathbb{R}$  \cite{Gubser:2016guj, Heydeman:2016ldy, Gubser:2017tsi, Gubser:2017qed, Qu:2018ned, Gubser:2018cha}. 
The most important feature is that in the correlation functions, all the distances are taken $p$-adic norms. The ultra-metric features of the $p$-adic field $\mathbb{Q}_p$ are automatically realized by the Bruhat-Tits tree. 
Finally we emphasize that in the tensor network realization of $p$-adic AdS/CFT, when computing the boundary correlation functions, the Witten diagrams emerge automatically in the bulk of the tensor network.

Another important aspect of $p$-adic AdS/CFT is the bulk reconstruction via the analogue of HKLL formula derived in \cite{Bhattacharyya:2017aly}, where it was also shown that, at the linear level, the bulk operator can be obtained by the $p$-adic wavelet transform of the boundary operator. 
For more detail and a realization of $p$-adic bulk reconstruction in terms of tensor network see \cite{Bhattacharyya:2017aly}.

In this paper, we will focus on the bulk computation of boundary correlation functions, which was reviewed in Section \ref{subsec: padicCFT}.
We will take a different route from earlier works, i.e.\ we will recover  correlation functions of the boundary $p$-adic CFT using Wilson line networks based on a Chern-Simons like formulation of the bulk theory.

\section{Towards a PGL$(2,\mathbb{Q}_p)$ Chern-Simons-eques theory}

The three-dimensional Einstein gravity has no propagating gravitational degrees of freedom in the bulk.  
Hence classically it is equivalent to a topological theory:\footnote{For quantum equivalence we need to specify the allowed configuration space on both sides, for details see \cite{Witten:2007kt}.} its action can be rewritten in terms of the 3D Chern-Simons theory with gauge group being the isometry of the spacetime \cite{Witten:2007kt}. 
For Lorentzian AdS$_3$, the isometry group is SL$(2,\mathbb{R})\times$SL$(2,\mathbb{R})$; for Euclidean AdS$_3$, it is SL$(2,\mathbb{C})$.

Many aspects of 3D gravity in asymptotically AdS$_3$ are more transparent in the corresponding Chern-Simons theory, especially the universal features of AdS$_3$/CFT$_2$.
For example, the appropriate asymptotic boundary condition in gravity can be translated into a gauge condition (e.g.\ Fefferman-Graham gauge) in the gauge theory side; the sources and vevs of the gravity side can be easily read off from the gauge field; the Virasoro Ward identity follows from the flatness condition of the gauge field, etc.

Therefore, to construct the bulk theory of the $p$-adic AdS/CFT, it is much easier to start with a gauge theory that is analogous to Chern-Simons theory. 
The Chern-Simons theory based on $\mathbb{R}$ or $\mathbb{C}$ can do much more than describing 3D gravity, e.g.\ it can reproduce knot invariants in compact three-manifolds. 
We will not attempt to construct the full $p$-adic Chern-Simons theory in this paper, but would only try to determine enough of its structures to show that its Wilson line network can reproduce the correlation functions of the boundary $p$-adic CFT. 

\subsection{Review: SL$(2)$ Chern-Simons in Asymptotic AdS$_3$}
\subsubsection{From 3D gravity to Chern-Simons theory}

The three-dimensional Einstein gravity with negative cosmological constant $\Lambda=-\frac{6}{\ell^2}$ has action
\begin{equation}\label{EHfirstorder}
S={1\over 4\pi G_N}\int_{\cal M} {\rm tr}\left(e\wedge R
+{1\over 3 \ell^2}e\wedge e\wedge e\right)
\end{equation}
where $R=d\omega+\omega\wedge \omega$ is the curvature two-form and $\omega$ the spin connection. 
Classically, the action is equivalent to the Chern-Simons theory \cite{Witten:1988hc, Witten:2007kt}:
\begin{equation}\label{EH-CS}
S=S_{\textrm{CS}}[A]-S_{\textrm{CS}}[\tilde A]+S_{\textrm{bndy}} \quad \textrm{with}\quad S_{\textrm{CS}}[A]
\equiv \frac{k}{4\pi}\int_{\mathcal{M}}\textrm{tr}(A\wedge dA 
+\frac{2}{3}A\wedge A \wedge A)
\end{equation}
with identification $k={\ell\over 4 G_N}$ and 
\begin{equation}\label{3beindef}
e={\ell\over2}\big(A-\tilde A\big) \qquad \textrm{and} \qquad  \omega={1\over 2}\big(A+\tilde A\big)
\end{equation}
For Lorentzian signature, $(A,\tilde{A})\in \mathfrak{sl}(2,\mathbb{R})$; whereas for Euclidean signature,
\begin{equation}
(A,\tilde{A})\in \mathfrak{sl}(2,\mathbb{C})\qquad \textrm{and}\qquad  \tilde{A}=-A^{\dagger}.
\end{equation}
We will focus on Euclidean signature. 

\subsection{PGL$(2,\mathbb{Q}_p)$ gauge theory on  Bruhat-Tits tree}

Now the task would be to construct a Chern-Simons-eques theory living on the Bruhat-Tits tree that can serve as a bulk gravity dual theory of the $p$-adic CFT.
Since the isometry group of the Bruhat-Tits tree is PGL$(2,\mathbb{Q}_p)$, this theory should be a lattice gauge theory of PGL$(2,\mathbb{Q}_p)$ living on the BT tree.

\subsubsection{PGL$(2,\mathbb{Q}_p)$ action on Bruhat-Tits tree}
\label{subsubsec:PGL2action}

Although the Bruhat-Tits tree is discrete, its isometry group is the continuous $\textrm{PGL}(2, \mathbb{Q}_p )$.
An element $\gamma\in \textrm{PGL}(2, \mathbb{Q}_p )$ acts on the lattice via
\begin{equation}
\langle \vec{f},\vec{g}\rangle  \rightarrow 
\langle\gamma \cdot \vec{f}, \gamma \cdot \vec{g} \rangle \ \qquad \textrm{with}\qquad \gamma=\left(\begin{matrix}a&b\\ c &d\end{matrix}\right) \in \textrm{PGL}(2,\mathbb{Q}_p).
\end{equation} 
Therefore it acts on a vertex on the Bruhat-Tits tree with coordinate (\ref{Hpcoord}) via
\begin{equation}
\begin{aligned}
\langle \langle 
\left(\begin{matrix}p^m\\ 0 \end{matrix}\right), \left(\begin{matrix}x^{(m)}\\ 1\end{matrix}\right)\rangle\rangle 
\qquad \longrightarrow \qquad 
\langle \langle \label{eq:treetrans}
\left(\begin{matrix} p^{m'}\\ 0 \end{matrix}\right), \left(\begin{matrix}\frac{a\,x^{(m)}+b}{c\,x^{(m)}+d}\\ 1\end{matrix}\right) 
\rangle\rangle 
\end{aligned}   
\end{equation}
where
\begin{equation}\label{padicSL2y}
p^{m'}= p^{m} \left|\frac{(c\,x+d)^2}{a d- b c} \right|_p   
\end{equation}
Namely, start with a bulk vertex $x=\sum^{m-1}_{ n=-N} a_n \,p^n$, with accuracy up to level $p^m$, its SL$(2,\mathbb{Q}_p)$ image is the bulk vertex
\begin{equation}
\frac{a\,x^{(m)}+b}{c\,x^{(m)}+d} =\sum^{m'-1}_{ n=-N} b_n \,p^n \qquad \quad \textrm{with accuracy } \quad p^{m'} .
\end{equation}

The PGL$(2,\mathbb{Q}_p)$ action (\ref{eq:treetrans}) with (\ref{padicSL2y}) suggests that the Bruhat-Tits tree is the analogue of the upper half plane (with Poincar\'e metric $ds^2=(dx^2+dy^2)/y^2$) for the $\mathbb{Q}_p$ line,  with $p^m$ playing the role of $y$-direction and $x^{(m)}$ the $x$-direction. This suggests that the cut-off surface in $p$-adic AdS/CFT should be the line of constant $p^N$, analogous to the choice of the $y=\epsilon$ surface as the cut-off surface for the AdS/CFT based on $\mathbb{R}$, where $y$ is the radial direction in Poincar\'e coordinates.

\subsubsection{Lattice gauge theory on Bruhat-Tits tree}

As in general formulation of lattice gauge theories, gauge connections are attached to links of the lattice. Namely, the gauge connection at each link takes values from the gauge group $G= \textrm{PGL}(2,\mathbb{Q}_p)$.
To be precise, the edges of the tree are labeled by
\begin{equation}
\textrm{edge:} \qquad (v,i) \qquad \textrm{with}\qquad i=-1, 0,1,\dots, p-1 
\end{equation}
and $i$ labels the $(p+1)$ edges starting from a vertex $v$. 
To each edge of the tree, we associate a connection
\begin{equation}
U(v, i) =U_{v\rightarrow v+\vec{i}}
\end{equation}
where $v+\vec{i}$ is $v$'s nearest neighbor along the direction-$i$. Note that we have also attached an orientation to the link. The same gauge connection with opposite choice of orientation of the given link is related by
\begin{equation}
U_{v\rightarrow v+\vec{i}} = U_{v+\vec{i} \rightarrow v}^{-1}
\end{equation}
Finally, we note that the link variable $U(v, i)$ in the lattice gauge theory is the direct analogue of the Wilson line in the continuous gauge theory:
\begin{equation}\label{linkWilson}
U(v,i) \sim W(v,i)= \exp(i \int_{\textrm{link}\, (v,i)} A)
\end{equation}

Under a gauge transformation, the connection in the specified orientation transforms as 
\begin{equation}
U_{v\rightarrow v+\vec{i}} \rightarrow g^{-1}(v) \, U_{v\rightarrow v+\vec{i}} \, g(v+\vec{i}) 
\end{equation}
where $g(v)$ is a PGL$(2,\mathbb{Q}_p)$ valued function on the vertex $v$ of the tree.
A pure gauge configuration is thus
\begin{equation} \label{puregauge}
U_{v\rightarrow v+\vec{i}}=g^{-1}(v) g(v+\vec{i})  \qquad \textrm{with}\qquad g(v)\in \textrm{PGL}(2,\mathbb{Q}_p).
\end{equation}
This is the analogue of the Wilson line segment (\ref{linkWilson}) with
\begin{equation}
A=g^{-1}dg \qquad \textrm{with} \qquad g \in \textrm{SL}(2,\mathbb{C}).
\end{equation}
for pure gauge configuration.

\subsection{Configuration space}
The Chern-Simons theory can reproduce almost all aspects of 3D Einstein gravity, except that at the level of path integral, we need to specify the allowed configuration space on both sides. (For more details, see \cite{Witten:2007kt}.)
For the gravity dual of $p$-adic CFT, this problem is more pronounced. As mentioned earlier, the fields of $p$-adic CFT can be chosen to be either complex-valued or $p$-adic valued. 
The $p$-adic CFT with complex-valued fields has a much smaller field space; in particular, all fields are conformal primaries --- there is no descendant! 
The $p$-adic CFT with $p$-adic valued fields has a much bigger field space and resembles ordinary 2D meromorphic CFT. 

What has mainly been studied in the literature is the $p$-adic CFT with complex valued fields, partially due to its simplicity.
In this paper, we would also like to first focus on this case, and try to construct a gravity dual (in terms of Chern-Simons like theory) of  $p$-adic CFT with complex valued fields. 

Since there is no descendent field in the boundary $p$-adic CFT, there is no stress-energy tensor. 
Accordingly, in its bulk dual, there is no bulk dynamics at all.
(We assume that BT tree has rigid length, i.e. the lengths of edges carry no information in the bulk theory.\footnote{It is possible to introduce bulk dynamics on the BT tree by allowing the lengths of edges to vary, see \cite{Gubser:2016htz}. However, we will not do so in this paper, since BT tree with rigid length is enough to capture bulk duals for all $p$-adic CFT with complex valued fields \cite{us2}. Nevertheless it would be interesting to study $p$-adic CFTs that are dual to theories living on the BT tree with varying edge lengths, or to other types of dynamical bulk theories. })
Note that for ordinary Einstein gravity in AdS$_3$, although there is no propagating gravitational degrees of freedom in the bulk, there are still boundary gravitons, sourced by the boundary stress-energy tensor. 
For the $p$-adic CFT with complex valued fields, there is not even boundary graviton. 
As a result, only one configuration is allowed: the one that is analogue to the pure AdS solution. 

\subsubsection{Review: Pure AdS$_3$ configuration}
The pure AdS$_3$ solution in Poincar\'e coordinates has metric
\begin{equation}\label{metricdefN}
ds^{2}_{\textrm{AdS}_3}=\frac{1}{\textrm{tr}\big[(L_{0})^2 \big]}
\textrm{tr}\big(e\otimes e\big)=\ell^2\frac{d\rho^2+dz d\bar{z}}{\rho^2}
\end{equation}
The corresponding gauge field configuration is a pure gauge with 
\begin{equation}
A=\mathfrak{g}^{-1}d\mathfrak{g} \qquad \textrm{with} \qquad \mathfrak{g}=e^{z L_{-1}}e^{\log{\rho}  \, L_{0}}
\end{equation}
and $\tilde{A}=-A^{\dagger}$. 
With the choice for the basis of the $\mathfrak{sl}(2)$ algebra
\begin{equation}
L_0=\begin{pmatrix}
\frac{1}{2} & 0\\
0& -\frac{1}{2}
\end{pmatrix}\qquad L_1=\begin{pmatrix}
0 & 0\\
-1& 0
\end{pmatrix}\qquad L_{-1}=\begin{pmatrix}
0 & 1\\
0& 0\end{pmatrix}
\end{equation}
we have explicitly
\begin{equation}\label{gEAdS}
\mathfrak{g}=\mathfrak{g}(\rho, z)=\frac{1}{\sqrt{\rho}}\begin{pmatrix}
\rho & z\\
0& 1\end{pmatrix} 
\end{equation}

The element $\mathfrak{g}$ can be used to parametrize the AdS$_3$ space. 
The Euclidean AdS$_3$ is a coset
\begin{equation}
\textrm{EAdS}_3=\frac{\textrm{SL}(2,\mathbb{C})}{\textrm{SU}(2)}
\end{equation}
The AdS$_3$ metric (\ref{metricdefN}) can be written as 
\begin{equation}
ds^{2}_{\textrm{AdS}_3}=\frac{\ell^2}{2}\textrm{tr}\big(G^{-1}dG G^{-1}dG\big)
\qquad \textrm{with}\qquad G=\mathfrak{g} \mathfrak{g}^{\dagger} 
\end{equation}
Note that $G$ is invariant under $\mathfrak{g}\mapsto \mathfrak{g} \cdot \gamma$ for $\forall \gamma \in \textrm{SU}(2)$ (i.e.\ the isotropy group of EAdS$_3$). 
Since we have used this gauge freedom to fix $\mathfrak{g}$ to the form in (\ref{gEAdS}), the points in EAdS$_3$ are mapped one-to-one to the corresponding element $\mathfrak{g}$:
\begin{equation}\label{EAdSmap}
(\rho, z,\bar{z}) \qquad \leftrightarrow \qquad \mathfrak{g}=\frac{1}{\sqrt{\rho}}\begin{pmatrix}
\rho & z\\
0& 1\end{pmatrix}
\end{equation}
In particular, the origin of AdS $(\rho=1, z=\bar z=0)$ is mapped to the identity matrix
\begin{equation}
\mathfrak{g}(o^{\textrm{AdS}_3})=\begin{pmatrix}
1& 0\\
0& 1\end{pmatrix} 
\end{equation}

\subsubsection{Vacuum solution of PGL$(2,\mathbb{Q}_p)$ Chern-Simons theory on BT tree}

As we have just reviewed, the gauge field configuration corresponding to pure AdS$_3$ can be constructed using the  SL$(2,\mathbb{C})/\textrm{SU}(2)$ elements (\ref{gEAdS}) that parametrize EAdS$_3$. 
For the Bruhat-Tits tree, we can similarly first parametrize it using PGL$(2,\mathbb{Q}_p)/\textrm{PGL}(2,\mathbb{Z}_p)$ elements, and then use them to build the PGL$(2,\mathbb{Q}_p)$ connections living on the edges of the tree.

Starting from the origin, a vertex $v$ on the Bruhat-Tits tree can be obtained by 
\begin{equation}
 \langle \langle \vec{f}_v, \vec{g}_v\rangle\rangle =\langle \langle \mathfrak{g}(v)\cdot \vec{f}_0, \mathfrak{g}(v)\cdot\vec{g}_0\rangle\rangle
\end{equation}
where the PGL$(2,\mathbb{Q}_p)$ element is
\begin{equation}\label{group0}
\mathfrak{g}(v) = \prod_{a_i\in \textrm{Path}[\mathfrak{o} \rightarrow v]} \Gamma[a_i\rightarrow a_{i+1}]
\end{equation}
where the Path$[\mathfrak{o}\rightarrow v]$ denotes the ordered set of vertices on the path from the origin $\mathfrak{o}$ to the vertex $v$, and the set of $\Gamma$'s are from the Hecke operator
\begin{equation}\label{Hecke}
\Gamma_{-1} \equiv \left(\begin{matrix}1 &0\\ 0&p \end{matrix}\right) \qquad \textrm{and} \qquad\Gamma_{n}\equiv \left(\begin{matrix}p &n\\ 0&1 \end{matrix}\right)\qquad n=0,1,\dots,p-1
\end{equation}
For example, the group element at the origin is just the identity:
\begin{equation}\label{OBT}
\mathfrak{g}(\mathfrak{o}^{BT})= \left(\begin{matrix}1 &0\\ 0&1 \end{matrix}\right)
\end{equation}
The points on the main branch all have
\begin{equation}
\mathfrak{g}(\textrm{main branch})= \left(\begin{matrix}p^n &0\\ 0&1 \end{matrix}\right) \qquad \textrm{with}\qquad n\in \mathbb{Z}
\end{equation}
More generally, we have a  map between a vertex on the tree to a PGL$(2,\mathbb{Q}_p)$ element 
\begin{equation}\label{vertextogroup}
\langle \langle \vec{f}, \vec{g}\rangle\rangle_{v} =\langle \langle \left(\begin{matrix}p^n\\ 0 \end{matrix}\right), \left(\begin{matrix}x^{(n)}\\ 1\end{matrix}\right)\rangle\rangle \qquad \longleftrightarrow \qquad \mathfrak{g}(v)= \begin{pmatrix}p^n & x^{(n)}\\ 0&1 \end{pmatrix} 
\end{equation}
The PGL$(2,\mathbb{Z}_p)$ redundancy (\ref{isotropy}) in the parameterization of the vertices is mapped to the gauge freedom $\mathfrak{g}(v)\mapsto  \mathfrak{g}(v)\cdot \gamma$ where $\gamma\in  \textrm{PGL}(2,\mathbb{Z}_p)$.
This is the direct analogue of the one-to-one map (\ref{EAdSmap}) for EAdS$_3$.
Finally, as explained in section \ref{subsubsec:PGL2action}, the cutoff surface (or line) that is suitable for $p$-adic AdS/CFT is the line of constant $p^N$, and $N$ should drop out at the end of the computation. 
Therefore, the group element for a boundary point $x$ is
\begin{equation}
\mathfrak{g}(\textrm{cutoff})= \left(\begin{matrix}p^N &x^{(N)}\\ 0&1 \end{matrix}\right) \qquad \textrm{with}\qquad N\in \mathbb{Z}
\end{equation}
where $x^{(N)}$ denotes $x$'s $p$-adic truncation at level $N$.

The vacuum solution of the  PGL$(2,\mathbb{Q}_p)$ Chern-Simons like theory is then just the pure gauge configuration (\ref{puregauge}) with $\mathfrak{g}(v)$ given by (\ref{vertextogroup}):
\begin{equation} \label{puregaugeEx}
U_{v\rightarrow v+\vec{i}}=\mathfrak{g}^{-1}(v) \mathfrak{g}(v+\vec{i})  \qquad \textrm{with}\qquad \mathfrak{g}(v)= \begin{pmatrix}p^n & x^{(n)}\\ 0&1 \end{pmatrix} 
\end{equation}
where $v+\vec{i}$ is $v$'s nearest neighbor along the direction-$i$, up to the gauge redundancy 
\begin{equation} \label{transform}
U_{v\rightarrow v+\vec{i}} \mapsto \gamma^{-1}_1 \cdot U_{v\rightarrow v+\vec{i}} \cdot \gamma_2
\end{equation}
where $\gamma_{1,2}\in  \textrm{PGL}(2,\mathbb{Z}_p)$.
Note that after we fix the asymptotically AdS boundary condition, this gauge freedom is reduced to 
\begin{equation}\label{gaugeR}
\gamma=\begin{pmatrix}a & b\\ 0&d \end{pmatrix} \in  \textrm{PGL}(2,\mathbb{Z}_p)\,.
\end{equation}

\subsection{PGL$(2,\mathbb{Q}_p)$ Wilson lines}

Having constructed the vacuum PGL$(2,\mathbb{Q}_p)$ configuration (\ref{puregaugeEx}) on the Bruhat-Tits tree, we now define Wilson lines on the tree, which serves as probes of the theory. 

\subsubsection{Wilson lines in fundamental representations}

As mentioned earlier, the link variable $U(v, i)$ in a lattice gauge theory plays the role of Wilson line from $v$ to $v+\vec{i}$ in the continuous theory (see (\ref{linkWilson})). 
A Wilson line from $v_1$ to $v_2$ is thus the ordered product of all the link variables on the path from $v_1$ to $v_2$:
\begin{equation}
\mathfrak{W}(v_1\to v_2) =\prod_{(v,i)\in \textrm{Path}[v_1\rightarrow v_2]}U(v,i)
\end{equation}

Now we focus on the PGL$(2,\mathbb{Q}_p)$  configuration (\ref{puregaugeEx}) that corresponds to the vacuum.
Its Wilson line $\mathfrak{W}^{0}$ only depends on the $\mathfrak{g}(v)$ at the initial and final points:
\begin{equation}\label{WLfund}
\mathfrak{W}^{0}_{\textrm{fund}}(v_1\to v_2) =\mathfrak{g}(v_1)^{-1}\, \mathfrak{g}(v_2) \qquad \textrm{with}\qquad \mathfrak{g}(v)= \begin{pmatrix}p^n & x^{(n)}\\ 0&1 \end{pmatrix} 
\end{equation}
where the subscript denotes the Wilson line in fundamental representation of  PGL$(2,\mathbb{Q}_p)$; since we will only consider Wilson lines for the vacuum configuration (\ref{puregaugeEx}), from now on we will drop the superscript ``0".
To reproduce correlation functions in $p$-adic CFT using Wilson lines, we need to first project the Wilson lines to representations that transform as conformal primaries. 

\subsubsection{PGL$(2,\mathbb{Q}_p)$ Representations for primaries}

First let's briefly review the case for 2D CFT based on $\mathbb{R}$.
We would only need to consider representations of the M\"obius symmetry SL$(2,\mathbb{C})$, since for $p$-adic CFT with complex-valued fields, the conformal symmetry is PGL$(2,\mathbb{Q}_p)$ --- there is no analogue of Virasoro symmetry. 

Recall that a unitary SL$(2,\mathbb{C})$ representation can be defined on the space of meromorphic functions. 
Under SL$(2,\mathbb{C})$, a primary field with conformal weight $h$ transforms as
\begin{equation} \label{MobiusC}
f_{h}(z)  \rightarrow \tilde{f}_{h}(\frac{az + b}{cz + d}) = \left(\frac{1}{(cz + d)^2}\right)^{-h}  f_{h}\left(z\right) \qquad  \textrm{with}\quad 
\begin{pmatrix}a & b\\ c&d \end{pmatrix} \in \textrm{SL}(2,\mathbb{C})
\end{equation}
One can also construct the representation of the primary field $f_{h}(z)$ using the generators of the Lie algebra $\mathfrak{sl}(2,\mathbb{C})$, which are 
\begin{equation}
L_{-1} = \partial_z, \qquad L_0 = z\partial_z + h, \qquad L_1 = \frac{1}{2} z^2\partial_z + h z. 
\end{equation}
in terms of $z$.
The resulting $\mathfrak{sl}(2,\mathbb{C})$ representation is discrete and spanned by its highest weight state $|h\rangle$ (defined by $L_{1} |h\rangle=0$ and $L_0 |h\rangle=h |h\rangle$) together with all its (global) descendants:
\begin{equation}\label{HWR}
|h\rangle \,, \quad L_{-1} |h\rangle \,, \quad (L_{-1})^2 |h\rangle \,, \quad \dots
\end{equation}

In 2D CFT based on $\mathbb{C}$, the second type of representation, i.e.\ the one constructed using the Lie algebra generators, is more familiar. 
However, as reviewed earlier, in a $p$-adic CFT with complex-valued fields, since all derivatives are zero, we cannot construct the second type of representation (\ref{HWR}). 
We can only define the highest weight states corresponding to primary fields; they will be used as projectors that relate the Wilson line operator in the fundamental representation (\ref{WLfund}) to correlation functions of primary fields.

The ``bra" and ``ket" states of a primary field $\mathcal{O}_{\Delta}$ of conformal weight $\Delta$ are defined as
\begin{equation}\label{Deltadef}
\langle \Delta |\equiv 
\langle \textrm{vac}| \mathcal{O}_{\Delta}(Z)
\qquad\textrm{and} \qquad 
|\Delta \rangle\equiv \mathcal{O}_{\Delta}(0) |\textrm{vac}\rangle 
\end{equation}
where the insertion point $Z$ is a free parameter for now.\footnote{In \cite{Fitzpatrick:2016mtp} $Z$ is taken to be $0$. We will see that different choices of $Z$ could lead to different results if the open ends of the Wilson lines are lying in the bulk. For special choices, the results would recover those prescribed in tensor network realizations of the $p$-adic AdS/CFT \cite{Bhattacharyya:2017aly, us2}. As we push the open ends of the Wilson lines towards the boundary, the dependence on $Z$ drops out. }
We have
\begin{equation}\label{normDelta}
\langle \Delta |\Delta \rangle=\frac{1}{|Z|^{2\Delta}}
\end{equation}
Both $\langle \Delta |$ and $|\Delta \rangle$ are invariant under the M\"obius transformation PGL$(2,\mathbb{Q}_p)$, i.e.\ they are merely states (which will serve as projectors) but not representations.

On the other hand, since  $\mathcal{O}_{\Delta}(x)$ transforms non-trivially under PGL$(2,\mathbb{Q}_p)$ (as (\ref{mobius})), there should exist another representation of $\mathcal{O}_{\Delta}(x)$ that characterizes this transformation property under PGL$(2,\mathbb{Q}_p)$. 
Let's first define the ``bra" state.
\begin{equation}\label{braX}
\langle X;\Delta |\equiv \langle \textrm{vac} |\mathcal{O}_{\Delta}(X)   \qquad \textrm{with} \qquad X \in \mathbb{Q}_p
\end{equation}
This is analogous to the meromorphic case in \cite{SimmonsDuffin:2012uy}, except that now $X\in \mathbb{Q}_p$ instead of $\in \mathbb{C}$. 
The set of states $\{\langle X ;\Delta |\}$ with $X \in \mathbb{Q}_p$ forms a representation of PGL$(2,\mathbb{Q}_p)$, transforming as 
\begin{equation}\label{Xmobius}
\langle X ;\Delta |\longrightarrow 
\left| \frac{ad-bc}{(cX+d)^2} \right|_p^{\Delta} {  \left\langle \frac{aX+b}{cX+d};\Delta \right| }\qquad \textrm{with} \quad \begin{pmatrix}
a& b\\
c& d
\end{pmatrix}\in \textrm{PGL}(2,\mathbb{Q}_p)
\end{equation}
Its projection to the primary state defined in (\ref{Deltadef}) is 
\begin{equation}\label{overlap1}
\langle X ;\Delta_i |\Delta_j\rangle =\frac{\delta_{ij}}{|X|_p^{2\Delta}}
\end{equation}

The construction of the corresponding ``ket" state is slightly more involved:
 \begin{equation}\label{ketX}
|X;\Delta \rangle\equiv \tilde{\mathcal{O}}_{\Delta}(X)  |\textrm{vac} \rangle
\end{equation}
where $\tilde{\mathcal{O}}_{\Delta}$ is the ``shadow operator"\footnote{For a review on shadow operators in CFTs based on $\mathbb{R}$, see e.g.\ \cite{SimmonsDuffin:2012uy}} of $\mathcal{O}_{\Delta}$, defined as
\begin{equation}\label{shadowdef}
\tilde{\mathcal{O}}_\Delta(X) \equiv \mathcal{N}(d,\Delta) \int_{\mathbb{Q}_p} dY |X-Y|_{p}^{2\Delta - 2d} \, \mathcal{O}_\Delta(Y)
\end{equation}
with 
\be \label{normN}
\mathcal{N}(d,\Delta) \equiv \frac{\zeta_p(2d-2\Delta)) \zeta_p(2\Delta)}{\zeta_p(d-2\Delta) \zeta_p(2\Delta-d)}
\ee
is the normalization constant and
$d$ is the dimension of the boundary theory, and we will mostly take $d=1$ in the rest of the paper unless otherwise stated.
The conformal dimension of the shadow operator follows from the definition (\ref{shadowdef}):
\begin{equation}\label{CDrelation}
\tilde{\Delta}_{\ell}=d-\Delta_{\ell}
\end{equation}
The most important property of the shadow operator $\tilde O_{\Delta}$ is the ``orthogonality" with the ordinary operator $\mathcal{O}_{\Delta}$:
\begin{equation} \label{OtO}
\begin{aligned}
 \langle \mathcal{O}_{\Delta}(Z) \tilde O_{\Delta}(Y)\rangle 
 = \delta(Z-Y) + \textrm{contact terms}, 
\end{aligned}
\end{equation}
We leave the proof of (\ref{OtO}) to Appendix (\ref{append:OtO}).
The overlap 
\begin{equation}\label{overlap2}
\langle \Delta_i |X;\Delta_j\rangle =\delta_{ij} \,
\delta(X-Z)
\end{equation}

For each $\mathcal{O}_{\Delta}$, there is a projection operator
\begin{equation}
\mathcal{P}_{\Delta}\equiv |X;\Delta \rangle \langle X;\Delta |
\end{equation}
which satisfies
\begin{equation}\label{projectorDelta}
\mathbf{1}_{\Delta}=\int_{\mathbb{Q}_p} dX |X;\Delta\rangle \langle X;\Delta |
\end{equation}
Finally, one can check that this is consistent with (\ref{normDelta}) and the two overlaps (\ref{overlap1}) and (\ref{overlap2}).

\subsubsection{Wilson lines in primary representations}

To obtain the Wilson line (from $v_1$ to $v_2$) in the primary representation of $\mathcal{O}_{\Delta}$, we use the projector (\ref{projectorDelta}) to project the Wilson line in fundamental representation (\ref{WLfund}) to the primary representation $|X;\Delta\rangle$:
\begin{equation} \label{Wilsonlinedef}
\hat{\mathfrak{W}}_\Delta(v_1\to v_2) 
=\int_{\mathbb{Q}_p} \,dX\, |X;\Delta\rangle  \hat{\mathfrak{W}}_{\textrm{fund}}(v_1\to v_2) \langle X;\Delta |
\end{equation}
Expressing $\mathfrak{W}_{\textrm{fund}}(v_1 \rightarrow v_2)$ in terms of the PGL$(2,\mathbb{Q}_p)$ element
\begin{equation}\label{WLfundexplicit}
\mathfrak{W}_{\textrm{fund}}(v_1\to v_2) =\begin{pmatrix}
a& b\\
c& d
\end{pmatrix}\in \textrm{PGL}(2,\mathbb{Q}_p)
\end{equation}
and then using the transformation property (\ref{Xmobius}), we obtain the Wilson line from $v_1$ to $v_2$ in representation $\Delta$:
\begin{equation}\label{WLprimarygeneric}
\hat{\mathfrak{W}}_\Delta(v_1\to v_2) 
=   \int_{\mathbb{Q}_p} \,dX\, \left| \frac{ad-bc}{(c X + d)^2}\right|_p^{\Delta}\, \left|X;\Delta\right\rangle \left\langle \frac{aX+b}{cX+d} ;\Delta \right|, 
\end{equation}
where $(a,b,c,d)$ is from (\ref{WLfundexplicit}).
Finally, from the orthogonality (\ref{overlap1}) and (\ref{overlap2}), the only nonzero component of (\ref{WLprimarygeneric}) is
\begin{equation}\label{WL2bulk}
\langle \Delta | \hat{\mathfrak{W}}_\Delta(v_1\to v_2) |\Delta \rangle=\frac{\left| (ad-bc) \right|^{\Delta}_p}{\left|a Z+b\right|^{2\Delta}_p} 
\end{equation}

\section{Correlation functions via Wilson line network}
\label{sec:WLN}

In the previous section we have constructed the Wilson lines for the PGL$(2,\mathbb{Q}_p)$ gauge theory living on the Bruhat-Tits tree. 
In particular, we have obtained the PGL$(2,\mathbb{Q}_p)$ connection that is analogous to the pure AdS$_3$ solution of the Chern-Simons theory.
In this section we will show that the bulk Wilson line networks built from these Wilson lines reproduce the correlation functions of the boundary $p$-adic CFT with complex-valued fields. 

\subsection{From bulk Wilson line segments to boundary two point functions}
\label{subsec:WLL2pt}

First let's consider the bulk Wilson line segment from $v_1$ to $v_2$, in the representation $\Delta $.
The Wilson line for the vacuum configuration (\ref{puregaugeEx}) and in the fundamental representation is
\begin{equation}\label{WLfundexpl}
\hat{\mathfrak{W}}_{\textrm{fund}}(v_1\rightarrow v_2)=\begin{pmatrix}
p^{n_2-n_1}& p^{-n_1}(x^{(n_2)}_2-x^{(n_1)}_1)\\
0& 1
\end{pmatrix}
\end{equation}
Plug this into (\ref{WL2bulk}) we have
\begin{equation}
\langle \Delta |\hat {\mathfrak{W}}_\Delta (v_1\to v_2) |\Delta \rangle = \frac{p^{-(n_1+n_2)\Delta}  
}{\left|p^{n_2} Z + (x^{(n_2)}_2-x^{(n_1)}_1)\right|_p^{2\Delta}}.
\end{equation}

\begin{figure}[h!]
        \centering
        \includegraphics[width=0.7\textwidth]{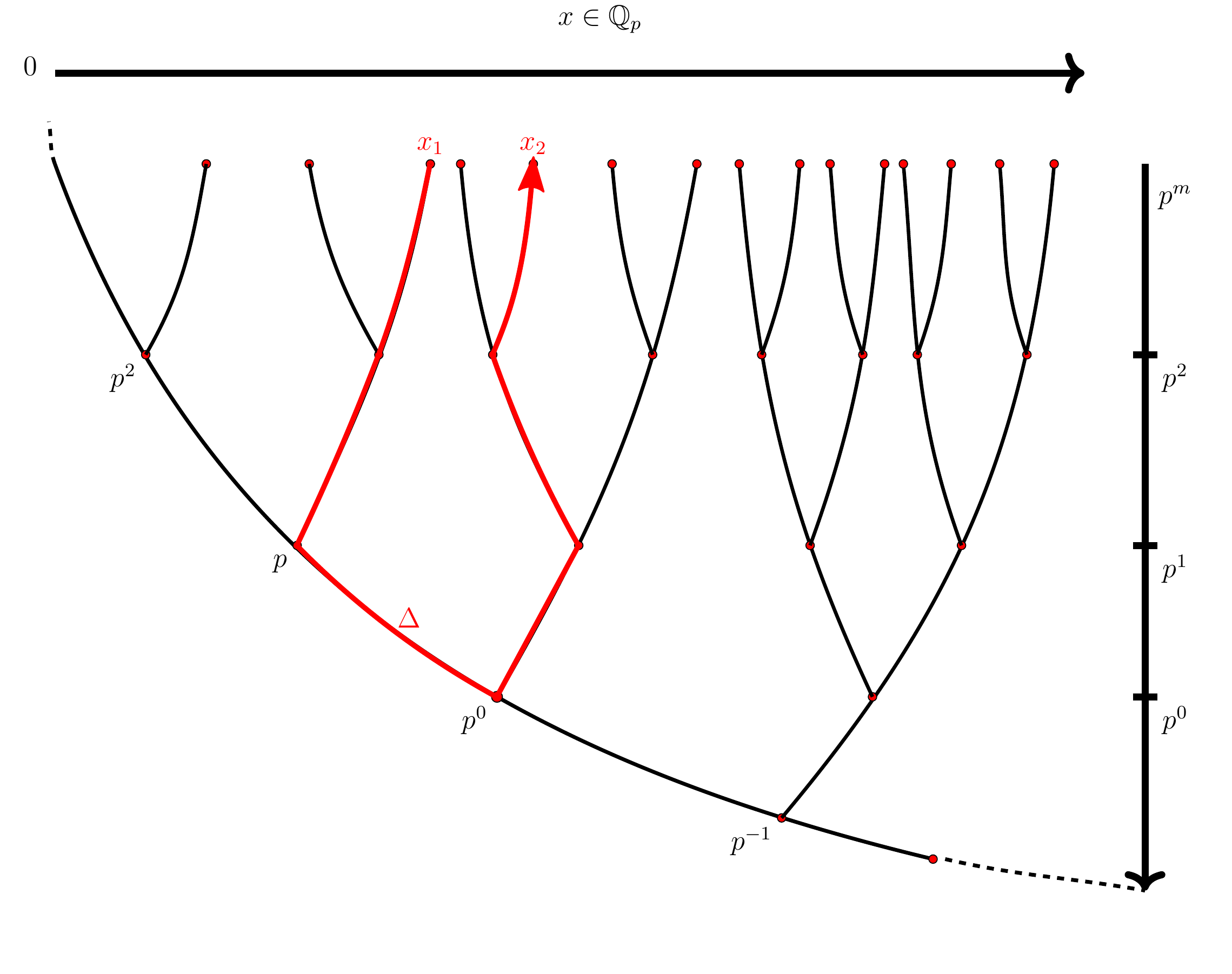}
        \caption{A Wilson line connecting two points at the boundary. }
        \label{fig:2ptWil}
\end{figure}

To reproduce the boundary two-point function, we push the two vertices to the boundary
\begin{equation}\label{bndy2point}
\langle \mathcal{O}_{\Delta}(x_1)\mathcal{O}_{\Delta}(x_2) \rangle= \lim_{v_i \rightarrow \textrm{bndy}}\langle \Delta | \hat{\mathfrak{W}}_{\Delta}(v_1 \rightarrow v_2) |\Delta \rangle
\end{equation}
(see Figure \ref{fig:2ptWil}.)
As $v_{i}$ approaches the boundary, $n_1=n_2\equiv N \to \infty$, the term $p^{n_2}Z$  in the $p$-adic norm in the denominator of (\ref{bndy2point}) drops out, since $|p^N Z|_p\rightarrow0$.
This gives
\begin{equation}
\begin{aligned}
\lim_{n_{1,2}=N\to \infty}\langle \Delta |\hat {\mathfrak{W}}_\Delta (v_1\to v_2) |\Delta \rangle &= \frac{p^{-2N\Delta}}{|x_2-x_1|_p^{2\Delta}}
\end{aligned}
\end{equation}
The gauge connection exchanges the auxiliary coordinate $X_i$ for an actual boundary coordinate $x_i$. 
Finally, in the overall factor $p^{-2N\Delta}$,  $N$ is the graph distance from the boundary points (on the cutoff surface), where the two ends of the Wilson line sit, to the origin of the tree $\mathfrak{o}^{\textrm{BT}}$.  
Therefore, the factor $p^{-2N\Delta}$ accounts for the radial dependence and can be absorbed into normalization of $\langle \Delta |$ and $|\Delta \rangle$.\footnote{
This is similar to the prescription in the tensor network construction \cite{Bhattacharyya:2017aly}. }
In summary, the Wilson line with two ends at the boundary correctly reproduces the two-point correlation function of the boundary $p$-adic CFT:
\begin{equation}
\begin{aligned}
\lim_{n_{1,2}=N\to \infty}\langle \Delta |\hat {\mathfrak{W}}_\Delta (v_1\to v_2) |\Delta \rangle =\langle \mathcal{O}_{\Delta}(x_1)\mathcal{O}_{\Delta}(x_2) \rangle
\end{aligned}
\end{equation}

We would like to pause here and discuss the residual gauge freedom after we fixed the boundary condition 
%at the boundary where 
at  $n_{1,2} \to \infty$. 
Following the standard treatment in the AdS/CFT correspondence, we should require the ``leading'' terms in the radial expansion of the fields be fixed, and the residue gauge transformation  $\gamma \in $ PGL($2, \mathbb{Q}_p$) should preserve these boundary conditions. 
%Following the same logic, it would suggest that the residual gauge transformation $\gamma \in $ PGL($2, \mathbb{Q}_p$) should be constrained. 
%Since we are not in a position to do a full  Brown-Henneaux asymptotic symmetry analysis, given that we do
Then the analogy with the asymptotic symmetry group analysis for asymptotic AdS$_3$ in the Chern-Simons formulation \cite{Banados:1998gg, Li:2015osa} might suggest that the residue gauge (i.e.\ the analogue of Penrose-Brown-Henneaux (PBH) transformation \cite{FG}) be 
\begin{equation}\label{RGguess}
\gamma=\begin{pmatrix}a & b\\ 0&a \end{pmatrix} \in  \textrm{PGL}(2,\mathbb{Q}_p)\,.
\end{equation}
However, since the boundary $p$-adic CFT is only sensitive to the ($p$-adic) norms, we only need to demand that the residue gauge transformation preserves the correlation functions.
As a result, the residue gauge transformation is bigger than (\ref{RGguess}):
\begin{equation}\label{residuegauge}
\gamma=\begin{pmatrix}a & b\\ 0&d \end{pmatrix} \in  \textrm{PGL}(2,\mathbb{Q}_p) \qquad \textrm{with} \qquad |a|_p=|d|_p 
\end{equation}
In particular, one can check that the gauge redundancy (\ref{gaugeR}) that arises from the isotropy group of $\mathbb{H}_p$ is inside this residue gauge (\ref{residuegauge}). 

\begin{comment}
 that determines the most sensible boundary expansion of $\gamma$,

 a natural requirement would be such that the correlation functions remain invariant under these residual transformations. 
Substituting (\ref{transform}) into (\ref{WL2bulk}) we conclude that the residual transformation should satisfy
\be
|a|_p=|d|_p  \qquad \textrm{and} \qquad c=0
\ee
near the boundary. 
\end{comment}

\subsection{From bulk Wilson line junctions to boundary three-point functions}
In the previous subsection we computed the expectation value of a single open Wilson line in our vacuum background (\ref{puregaugeEx}), and showed that when the vertices are pushed to the boundary it reproduces the two-point function of the $p$-adic CFT. 
Now we move on to the three-point function. 

\subsubsection{Bulk three-way junction}
\label{subsubsec:bulkWLJ}

\begin{figure}[h!]
        \centering
        \includegraphics[width=0.7\textwidth]{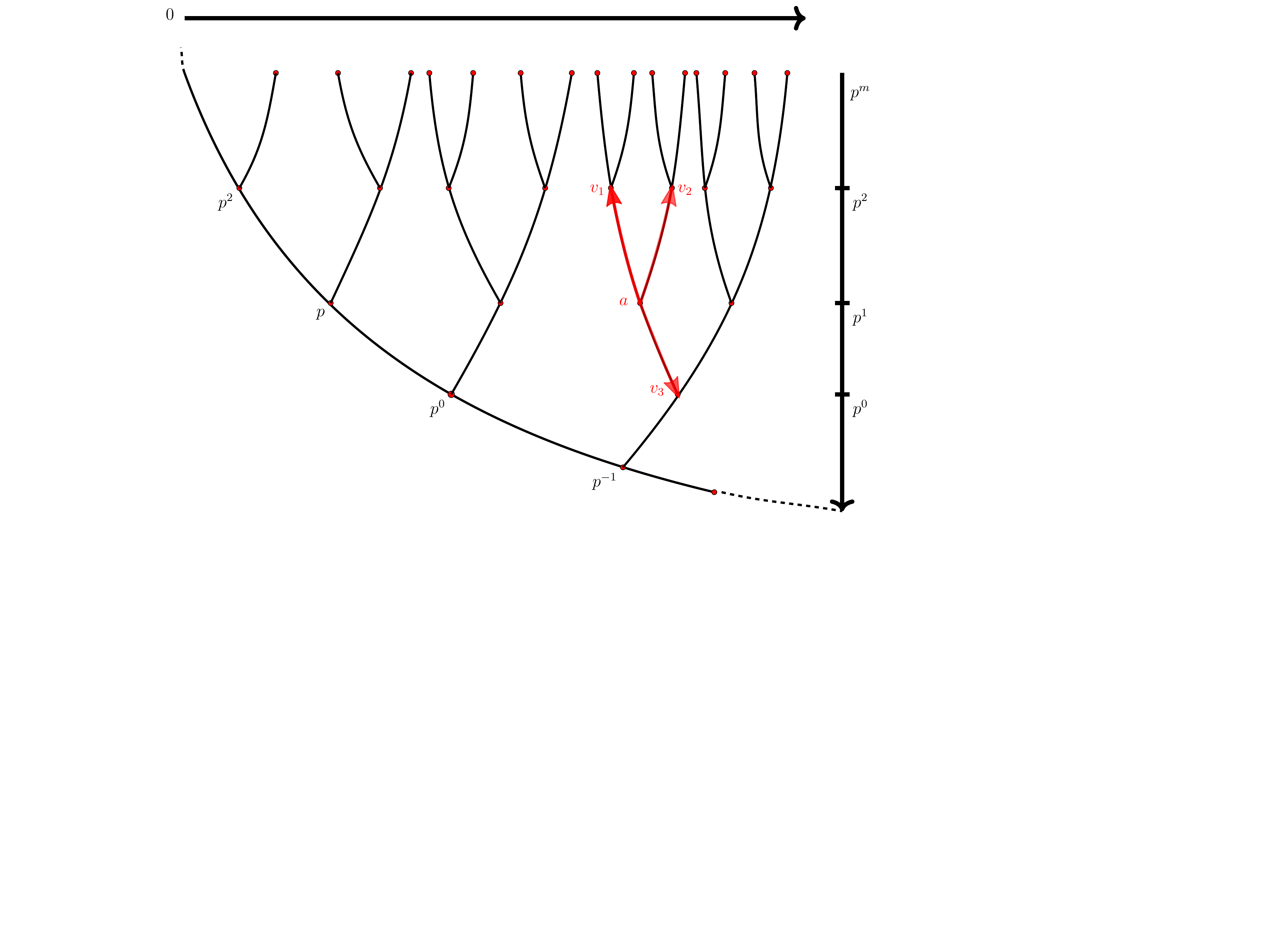}
        \caption{A junction of three Wilson lines connecting its nearest neighbours.  }
        \label{fig:p2neighbour}
\end{figure}

The relevant bulk object is the three-way junction of Wilson lines,   see Figure \ref{fig:p2neighbour} for the case of $p=2$.
Consider the product of three Wilson lines, all starting at the vertex $v_a$ on the Bruhat-Tits tree but ending at the vertices $v_{1,2,3}$
and carrying representation $\Delta_{1,2,3}$, respectively:
\begin{equation}\label{3wayjunction}
\prod_{i=1}^{3}\hat {\mathfrak{W}}_{\Delta_i} (v_a\to v_i )  =\left[ \prod^{3}_{i=1} \int_{\mathbb{Q}_p} \,dX_i\, |X_i ;\Delta_i\rangle \right]\left[\prod^3_{i=1}  \hat{\mathfrak{W}}_{\textrm{fund}}(v_a\to v_i) \right]\left[\prod^3_{i=1} \langle X_i;\Delta_i |\right]
\end{equation}
The three end points $v_i$ should be contracted with $\langle \Delta_i |$. 
At the vertex $v_a$, where the Wilson lines with three different representations meet, the bulk gauge invariance requires that the product of $\langle X_i;\Delta_i |$ be projected to the singlet $|\mathcal{S}\rangle$ that arises from the tensor product of the three representations.
Namely, our final expression should be  
\begin{equation}
\left[\prod^3_{i=1} \langle \Delta_i |\right]  \left[\prod^3_{i=1} \hat {\mathfrak{W}}_{\Delta_i} (v_a\to v_i)  \right]|\mathcal{S}\rangle
\end{equation}
which can be broken into three parts:
\begin{equation}\label{3wayjunctionfinal}
\left[ \prod^3_{i=1} \int \,dX_i\,  \langle \Delta_i |X_i ;\Delta_i\rangle \right] \,,\qquad \left[\prod^3_{i=1}  \hat{\mathfrak{W}}_{\textrm{fund}}(v_a\to v_i) \right]\,,\qquad \left[\prod^3_{i=1} \langle X_i;\Delta_i |\right]|\mathcal{S}\rangle
\end{equation}

As before, let's consider the three terms starting from the right. First
\begin{equation}
\begin{aligned}
\left[\prod^3_{i=1}  \langle X_i;\Delta_i | \right] |\mathcal{S}\rangle
&\equiv P_{123}(X_1,X_2,X_3)
\end{aligned}
\end{equation}
where $P$ is the intertwiners of the group and is fixed by the PGL$(2,\mathbb{Q}_p)$ invariance to be
\begin{equation}\label{intertwiner}
\begin{aligned}
P_{123}(X_1,X_2,X_3)=\frac{C_{123}}{|X_{12}|_p^{\Delta_1+\Delta_2-\Delta_3}|X_{13}|_p^{\Delta_1+\Delta_3-\Delta_2}|X_{23}|_p^{\Delta_2+\Delta_3-\Delta_1}}
\end{aligned}
\end{equation}
where the structure constants $C_{123}$ come from  the defining data of the dual $p$-adic CFT. 
This is completely parallel to the SL$(2,\mathbb{C})$ case in \cite{Fitzpatrick:2016mtp}.

Next consider the middle bracket in (\ref{3wayjunctionfinal}): the product of the three Wilson lines. Each one is given by
\begin{equation}
\hat{\mathfrak{W}}_{\textrm{fund}}(v_a\rightarrow v_i)=\begin{pmatrix}
p^{n_i-n_a}& p^{-n_a}(x^{(n_i)}_i-x_{a})\\
0& 1
\end{pmatrix}\in \textrm{PGL}(2,\mathbb{Q}_p)
\end{equation}
and acts on $\langle X_i ;\Delta_i|$ via (\ref{Xmobius}).
Since the intertwiner $P$ in (\ref{intertwiner})  is constructed by the product of $\langle X_i ;\Delta_i|$, the action of the product of these three Wilson lines on the intertwiner is
\begin{equation} \label{transP}
\begin{aligned}
&\left[ \prod_{i=1}^3  \hat{\mathfrak{W}}_{\textrm{fund}}(v_a\rightarrow v_i)\right]  P_{123}(X_1,X_2,X_3)= \left[ \prod_i  p^{(n_a-n_i)\Delta_i} \right]
P_{123}\left(X'_1,X'_2, X'_3\right),
\end{aligned}
\end{equation}
where
\begin{equation}
X'_{i}\equiv p^{n_i-n_a} X_i +p^{-n_a}(x^{(n_i)}_i-x^{}_a)\end{equation}

The last step is to contract (\ref{transP}) with the three $\langle \Delta_i |$.
Using (\ref{overlap2}), we have
\begin{equation} \label{3pt}
\begin{aligned}
&\left[\prod^3_{i=1} \langle \Delta_i |\right]   \hat {\mathfrak{W}}_{\Delta_i} (v_a\to v_i)  |\mathcal{S}\rangle\\
& =  \frac{p^{-n_1\Delta_1 - n_2\Delta_2-n_3\Delta_3} {C_{123} }}{|(p^{n_1}-p^{n_2})Z+ x_{12}|_p^{\Delta_1+\Delta_2-\Delta_3} |(p^{n_2}-p^{n_3})Z+ x_{23}|_p^{\Delta_2+\Delta_3-\Delta_1}|(p^{n_3}-p^{n_1})Z+ x_{31}|_p^{\Delta_3+\Delta_1-\Delta_2}}, 
\end{aligned}
\end{equation}
where $x_{ij} \equiv x^{(n_i)}_i- x^{(n_j)}_j$.

\subsubsection{Boundary three-point function}
\label{subsubsec:bndy3pt}

\begin{figure}[h!]
        \centering
        \includegraphics[width=0.7\textwidth]{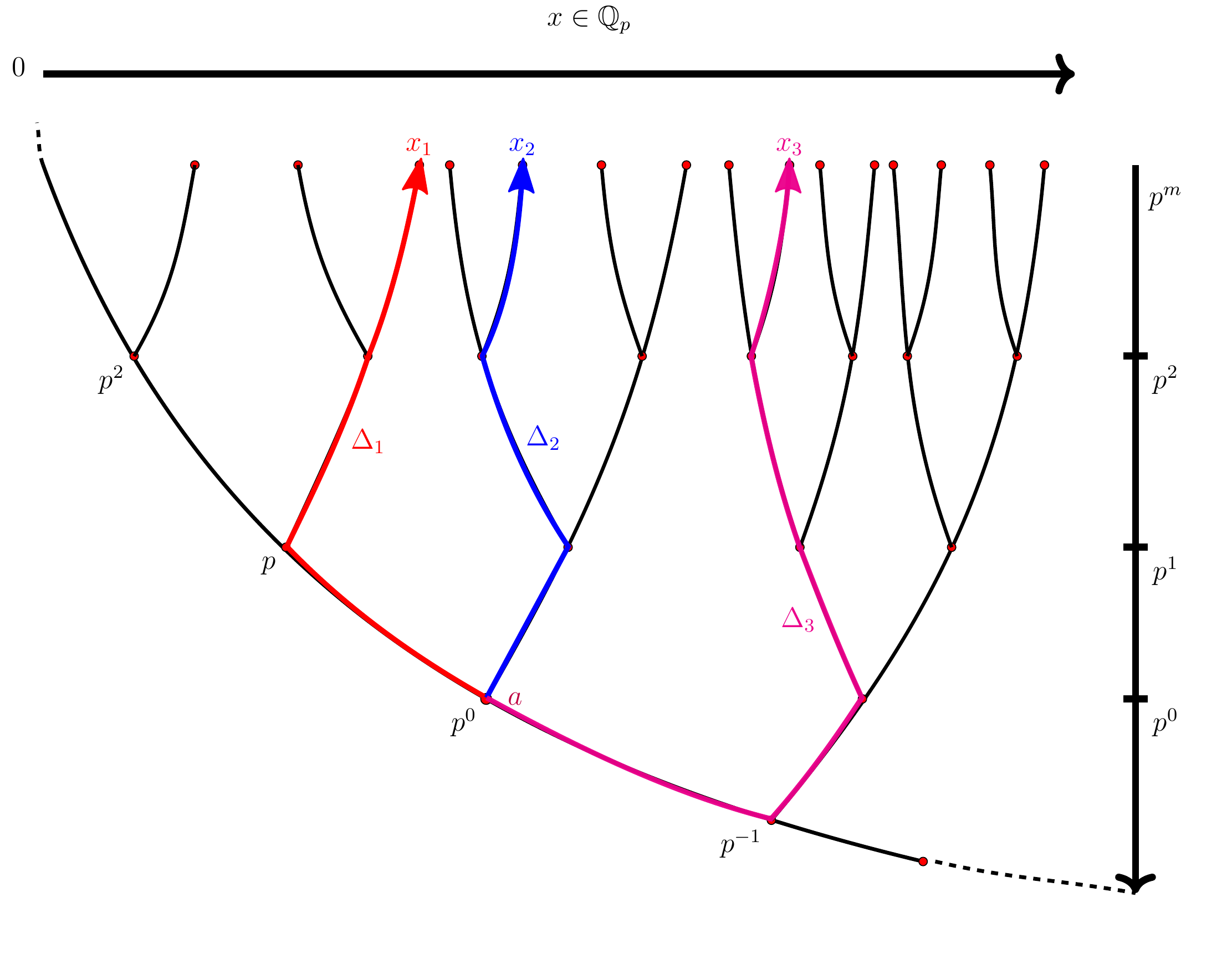}
        \caption{A network of three Wilson lines connecting three boundary points $x_{1,2,3}$ and joining at a junction $a$. }
        \label{fig:3pt}
\end{figure}

As in the two-point function, to reproduce the three-point correlation function in $p$-adic CFT, we push the three vertices $v_{i}$
in the three-way Wilson line junction (\ref{3pt}) to the boundary,  by taking the limit 
$n_{1,2,3}\equiv N \to \infty$. 
(See Figure \ref{fig:3pt}.)
Again, since $|p^NZ|_p\rightarrow 0$, $Z$ drops out in this limit, and the three-point Wilson line network reduces to
\begin{equation}
\begin{aligned}
\lim_{N\to \infty}\left[\prod^3_{i=1} \langle \Delta_i |\right]   \hat {\mathfrak{W}}_{\Delta_i} (v_a\to v_i)  |\mathcal{S}\rangle & = \frac{p^{-N ( \Delta_1 +\Delta_2+\Delta_3)} {C_{123} }}{| x_{12}|_p^{\Delta_1+\Delta_2-\Delta_3} |x_{23}|_p^{\Delta_2+\Delta_3-\Delta_1}|x_{31}|_p^{\Delta_3+\Delta_1-\Delta_2}}\\
\end{aligned}
\end{equation}
We see that the overall normalization consists of factors $p^{-N \Delta_i}$, where $N$ is the graph distance between boundary points (on the cutoff surface) and the origin of the tree $\mathfrak{o}^{\textrm{BT}}$. As in matching the expectation value of the Wilson line segment with the boundary two-point function in Section.~\ref{subsec:WLL2pt}, these  factors are again absorbed into normalization of $\langle \Delta_i |$ and $|\Delta_i \rangle$.
To summarize, the trivalent Wilson line network with endpoint on the boundary reproduces the three-point correlation function of the boundary $p$-adic CFT:
\begin{equation}
\begin{aligned}
\lim_{N\to \infty}\left[\prod^3_{i=1} \langle \Delta_i |\right]   \hat {\mathfrak{W}}_{\Delta_i} (v_a\to v_i)  |\mathcal{S}\rangle 
=\langle \mathcal{O}_{1}(x_1)\mathcal{O}_{2}(x_2) \mathcal{O}_{3}(x_3) \rangle
\end{aligned}
\end{equation}
Finally, we mention that the final result does not depend on the position of the internal vertex $v_a$, due to the fact that the configuration (\ref{puregaugeEx}) is a pure gauge.
This is completely analogous to the $\mathbb{R}$ case.

\subsection{Wilson line network and four-point functions}

Finally we consider the Wilson line network that reproduces the boundary four-point functions. 
The main input is the three-point intertwiner (\ref{intertwiner}).

\subsubsection{Bulk Wilson line network}

Given the positions $x_i$ with $i=1$ to $4$ on the boundary, there are different ways they can join in the bulk, depending on relative distances between them. 
Recall that every triangle in $p$-adic fields is an isosceles triangle whose legs are longer than its base. 
It follows from similar arguments that for any four points $x_i$ with $i=1$ to $4$, there are only two possible configurations for the six distances: 
\begin{itemize}
\item $(1,2,3)$\\
\begin{equation}
\textrm{WLOG:}\qquad |x_{12}|_p \leq |x_{13}|_p=|x_{23}|_p \leq |x_{14}|_p=|x_{24}|_p=|x_{34}|_p
\end{equation} 
\item $(1,1,4)$
\begin{equation}
\textrm{WLOG:}\qquad|x_{12}|_p\,, \,\,\, |x_{34}|_p\leq |x_{13}|_p =|x_{14}|_p=|x_{23}|_p=|x_{24}|_p
\end{equation} 
\end{itemize}
To compare with the four-point functions in (\ref{4ptpadic}), let's focus on the first case.\footnote{The second case can be either treated explicitly in the same manner or mapped to the first case via a $p$-adic M\"obius transformation.} WLOG, we can choose the $|x_i|_p\leq |x_4|_p$. Then the $(s, t, u)$ channels are 
\begin{eqnarray}
\textrm{s-channel:}&\quad&|x_{12}|_p \, <\,|x_{13}|_p=|x_{23}|_p \,< \,|x_{14}|_p=|x_{24}|_p=|x_{34}|_p \label{schannel}\\
\textrm{t-channel:}&\qquad&|x_{23}|_p \, <\,|x_{12}|_p=|x_{13}|_p\, <\, |x_{14}|_p=|x_{24}|_p=|x_{34}|_p\label{tchannel}\\
\textrm{u-channel:}&\qquad&|x_{13}|_p \, <\, |x_{12}|_p=|x_{23}|_p \, <\, |x_{14}|_p=|x_{24}|_p=|x_{34}|_p \label{uchannel}
\end{eqnarray}
Now we will focus on the four-point function in its s-channel and construct a bulk Wilson line network to reproduce it explicitly; the other two cases are completely parallel. 

\begin{figure}[h!]
        \centering
        \includegraphics[width=0.7\textwidth]{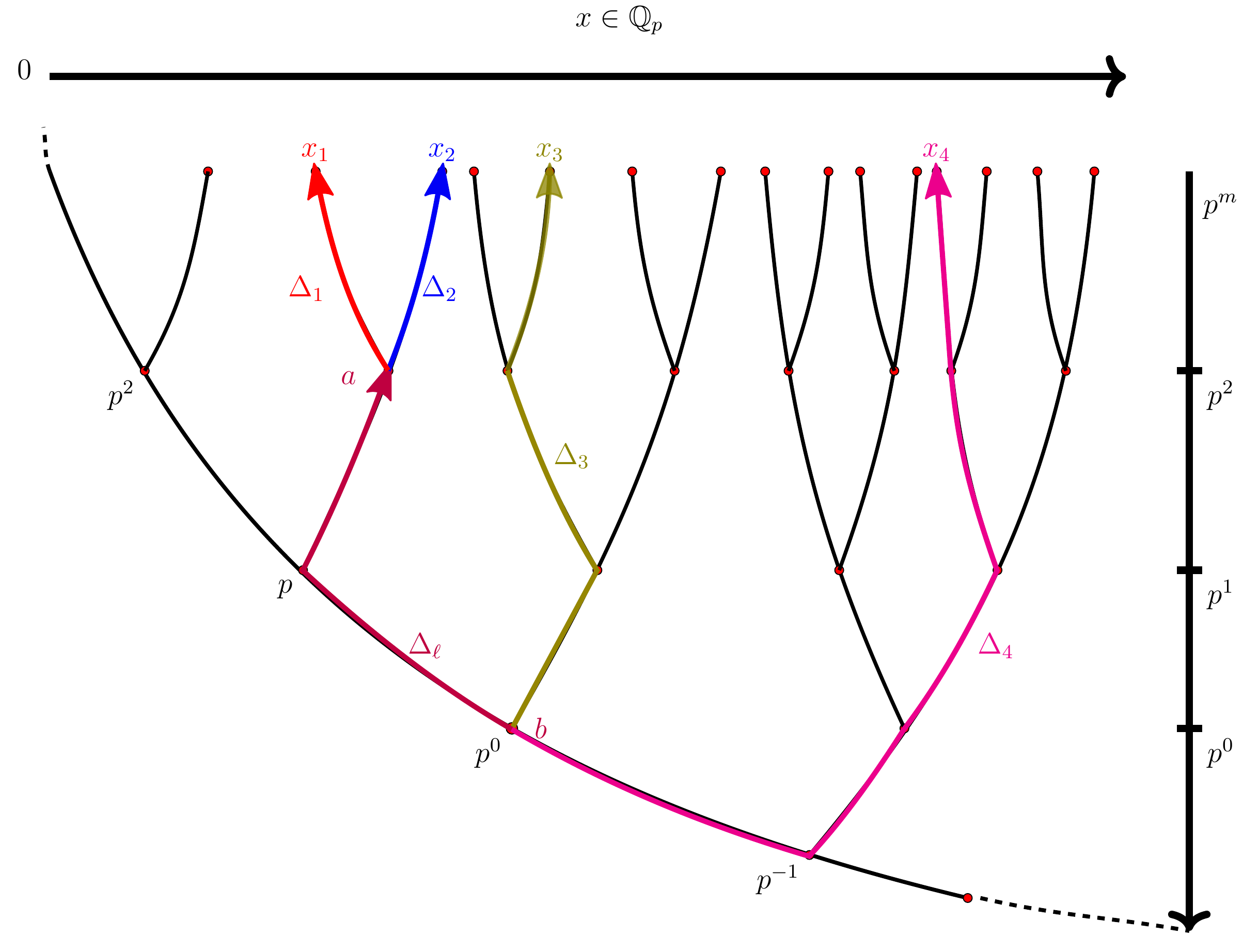}
        \caption{A network of four Wilson lines with two intermediate junctions. }
        \label{fig:4pt}
\end{figure}

Take a pair of three-way Wilson line junctions considered in the previous subsection, and join the two to form the following Wilson line network, illustrated in figure \ref{fig:4pt}.
There are four external vertices $v_i$ with $i=1,2,3,4$ and two internal vertices $v_a$ and $v_b$; and correspondingly four external legs and one internal leg from $v_b$ to $v_a$. 
The configuration in the bulk corresponds to the s-channel:
\begin{equation}\label{crab}
\left[\prod^{2}_{i=1}\hat {\mathfrak{W}}_{\Delta_i} (v_a\to v_i)  \right] 
\left[\sum_{\ell}\hat {\mathfrak{W}}_{\Delta_{\ell}} (v_b\to v_a)\right]
\left[\prod^{4}_{i=3}\hat {\mathfrak{W}}_{\Delta_i} (v_b\to v_i)  \right] 
\end{equation}
where $\ell$ runs over all the primary fields $\mathcal{O}_{\ell}$ that appear in the OPE channel of $\mathcal{O}_1$ and $\mathcal{O}_2$.
As before, we first project to the continuous basis: the four external legs with $X_i$ basis and the internal leg with $Y$ basis: 
\begin{equation}
\begin{aligned}
\left[\prod^{2}_{i=1}\hat {\mathfrak{W}}_{\Delta_i} (v_a\to v_i)  \right] &= \left[ \prod^{2}_{i=1} \int_{\mathbb{Q}_p} \,dX_i\, |X_i ;\Delta_i\rangle \right]\left[\prod^2_{i=1}  \hat{\mathfrak{W}}_{\textrm{fund}}(v_a\to v_i) \right]\left[\prod^2_{i=1} \langle X_i;\Delta_i |\right]\\
\sum_{\ell}\hat {\mathfrak{W}}_{\Delta_{\ell}} (v_b\to v_a)&=  \sum_{\ell}\int_{\mathbb{Q}_p} \,dY\, |Y;\Delta_{\ell}\rangle \,\hat{\mathfrak{W}}_{\textrm{fund}}(v_b\to v_a) \,\langle Y;\Delta_{\ell} |\\
\left[\prod^{4}_{i=3}\hat {\mathfrak{W}}_{\Delta_i} (v_b\to v_i)  \right] &= \left[ \prod^{4}_{i=3} \int_{\mathbb{Q}_p} \,dX_i\, |X_i ;\Delta_i\rangle \right]\left[\prod^4_{i=3}  \hat{\mathfrak{W}}_{\textrm{fund}}(v_b\to v_i) \right]\left[\prod^4_{i=3} \langle X_i;\Delta_i |\right]
\end{aligned}
\end{equation}
In these basis, the Wilson line network (\ref{crab}) contains  five ``bra" states and five ``ket" states in total. 
Let's first determine how to contract to obtain the final observable.

As before, the four external ``bra" states $|X_i;\Delta_i\rangle$ should be contracted with the corresponding $\langle \Delta_i |$.
The remaining four $\langle X_i;\Delta_i |$ and one $\langle Y;\Delta_{\ell}|$ and one $|Y;\Delta_{\ell}\rangle$ are contracted as follows.
Consider the junction at vertex $v_b$ first. 
As explained before, the bulk gauge invariance demands that the tensor product of these three representations be contracted with the singlet:
\begin{equation}
\begin{aligned}
\langle Y;\Delta_{\ell} |
\langle X_3;\Delta_{3}|
\langle X_4;\Delta_{4}
|\mathcal{S}\rangle&=P_{\ell 3 4}(Y,X_3,X_4)\\
&\equiv \langle \mathcal{O}_{\ell}(Y) \mathcal{O}_{3}(X_3) \mathcal{O}_{4}(X_4)  \rangle
\end{aligned}
\end{equation}
The remaining $\langle X_1;\Delta_1 |$, $\langle X_2;\Delta_2 |$, and $|Y;\Delta_{\ell}\rangle$ form
\begin{equation}
\begin{aligned}
\langle X_1;\Delta_{1}|\langle X_2;\Delta_{2}| Y;\Delta_{\ell}\rangle&=\langle \mathcal{O}_{1}(X_1) \mathcal{O}_{2}(X_2) \tilde{\mathcal{O}}_{\ell}(Y) \rangle\\&=P_{12 \tilde{\ell}}(X_1,X_2,Y)
\end{aligned}
\end{equation}
where we have used the definition of the ``bra" state
\begin{equation} 
|Y;\Delta_{\ell}\rangle \equiv \tilde{\mathcal{O}}_{\ell}(Y) |\textrm{vac}\rangle 
\end{equation}
where $\tilde{\mathcal{O}}_{\ell}$ is the shadow of $\mathcal{O}_{\ell}$:
\begin{equation}
\tilde{\mathcal{O}}_{\ell}(Y) \equiv \mathcal{N}(1,\Delta_{\ell}) \int_{\mathbb{Q}_p} dU \, |Y-U|_{p}^{2\Delta_{\ell} - 2}\, \mathcal{O}_{\ell}(U)
\end{equation}

With the prescription of the contraction above, the expectation value of the Wilson line network (\ref{crab}) is
\begin{equation}
\begin{aligned}
W_4\equiv&\left[\prod^4_{i=1} \langle \Delta_i |\right] \left[\prod^{2}_{i=1}\hat {\mathfrak{W}}_{\Delta_i} (v_a\to v_i)  \right] 
\hat {\mathfrak{W}}_{\Delta_{\ell}} (v_b\to v_a)
\left[\prod^{4}_{i=3}\hat {\mathfrak{W}}_{\Delta_i} (v_b\to v_i)  \right] |\mathcal{S}\rangle\\
=&\left[ \prod^4_{i=1} \int_{\mathbb{Q}_p} \,dX_i\,  \langle \Delta_i |X_i ;\Delta_i\rangle \right]\int_{\mathbb{Q}_p} dY 
\left[\prod^{2}_{i=1}\hat {\mathfrak{W}}_{\textrm{fund}} (v_a\to v_i)  \right]  P_{12 \tilde{\ell}}(X_1,X_2,Y)\\
& \qquad \qquad \qquad  \,\,\,\,\,\quad\hat {\mathfrak{W}}_{\textrm{fund}} (v_b\rightarrow v_a) \left[\prod^{4}_{i=3}\hat {\mathfrak{W}}_{\textrm{fund}} (v_b\rightarrow v_i)  \right] P_{\ell34}(Y,X_3,X_4)
\end{aligned}
\end{equation}
where $\hat{\mathfrak{W}}_{\textrm{fund}}$ are the Wilson line (in fundamental representation) segment (see \ (\ref{WLfundexpl})) in the vacuum background (\ref{puregaugeEx}).
Applying the ``$\hat{\mathfrak{W}}$" on the matching ``bra" states $\langle X_i;\Delta_i|$, using the overlap  (\ref{overlap2}), and finally evaluating the $\int_{\mathbb{Q}_p} dX_i$ integrals  we have
\begin{equation}\label{bulkcrab}
\begin{aligned}
W_4=&c
\sum_{\ell}\int_{\mathbb{Q}_p} dY P_{1 2 \tilde{\ell}}(X'_1,X'_2,Y) P_{\ell 34}(Y',X'_3,X'_4)
\end{aligned}
\end{equation}
with
\begin{equation}
\begin{aligned}
X'_i&\equiv\begin{cases} &p^{n_i-n_a}Z+p^{-n_a}(x_i-x_a) \quad \textrm{for } i=1,2\\
&p^{n_i-n_b}Z+p^{-n_b}(x_i-x_b) \quad \,\textrm{for } i=3,4,
\end{cases}\\
Y'&\equiv\qquad p^{n_a-n_b}Y+p^{-n_b}(x_a-x_b)\\
\end{aligned}
\end{equation}
and $c$ a constant:
\begin{equation}
c=p^{-\sum^4_{i=1}n_i\Delta_i} p^{n_a (\Delta_1+\Delta_2-\Delta_{\ell})}p^{n_b (\Delta_3+\Delta_4+\Delta_{\ell})}
\end{equation}

\subsubsection{Boundary four-point function and conformal blocks}
Now we push the four external vertices $v_i$ with $i=1,2,3,4$ to the boundary, namely, in (\ref{bulkcrab}) we take the limit 
\begin{equation}
n_i=N\rightarrow \infty
\end{equation}
In this limit, the  parameter $Z$ again drops out, and we get
\begin{equation}\label{W4integral}
\begin{aligned}
W_4=&p^{-N\sum^4_{i=1}\Delta_i}\sum_{\ell}\int_{\mathbb{Q}_p} dy \, P_{12 \tilde{\ell}}(x_1,x_2,y) \,P_{\ell34}(y, x_3, x_4)
\end{aligned}
\end{equation}
where we have used the relation (\ref{CDrelation}) between the conformal dimensions of an operator and its shadow, and changed the integration variable from $Y$ to $y\equiv p^{n_a}Y+x_a$ (which implies $dY=p^{n_a} dy$).
The three-point function is
\begin{equation}\label{3ptO34}
P_{\ell 34}(y, x_3, x_4)=\frac{C_{\ell 3 4}}{\left|y-x_3\right|_p^{\Delta_{\ell}+\Delta_3-\Delta_{4}} \left|x_3-x_{4}\right|_p^{\Delta_3+\Delta_{4}-\Delta_{\ell}}\left|x_{4}-y\right|_p^{\Delta_{4}+\Delta_{\ell}-\Delta_{3}} }
\end{equation}
And the three-point function involving a shadow operator is
\begin{equation}\label{3pt12shadow}
\begin{aligned}
P_{12 \tilde{\ell}}(x_1,x_2,y)& 
\equiv \langle \mathcal{O}_{1}(x_1) \mathcal{O}_{2}(x_2) \tilde{\mathcal{O}}_{\ell}(y) \rangle \\
& = \sum_{\ell'}\frac{C_{12\ell'}}{|x_{12}|^{\Delta_1+\Delta_2-\Delta_{\ell'}}} \langle \mathcal{O}_{\ell'}(x_2) \tilde{\mathcal{O}}_{\ell}(y) \rangle  \\
&= \frac{C_{12\ell}}{|x_{12}|^{\Delta_1+\Delta_2-\Delta_{\ell}}} \delta(x_2-y) 
\end{aligned}
\end{equation}
where we have first evaluated the OPE between $\mathcal{O}_1$ and $\mathcal{O}_2$ and then used the orthogonality condition (\ref{OtO}) between an operator and its shadow.\footnote{For an alternative derivation see Appendix.~\ref{append:3ptshadow}.}
Plugging (\ref{3ptO34}) and (\ref{3pt12shadow}) into the integral (\ref{W4integral}) we get
\begin{equation}\label{W4final}
\begin{aligned}
W_4=&p^{-N\sum^4_{i=1}\Delta_i}\sum_{\ell}\frac{C_{12\ell}C_{\ell 34}}{|x_{12}|_p^{\Delta_1+\Delta_2-\Delta_{\ell}}|x_{23}|_p^{\Delta_{\ell}+\Delta_3-\Delta_{4}}|x_{34}|_p^{\Delta_3+\Delta_4-\Delta_{\ell}}|x_{24}|_p^{\Delta_4+\Delta_{\ell}-\Delta_{3}}}
\end{aligned}
\end{equation}
where the factor $p^{-N\sum^4_{i=1}\Delta_i}$ can be absorbed into the normalization of  $\langle \Delta_i |$ and $|\Delta_i \rangle$, consistent with the prescription for two-point functions in Section.~\ref{subsec:WLL2pt} and three-point functions in Section.~\ref{subsubsec:bndy3pt}.

Finally, since this is the s-channel configuration, we plug in the configuration (\ref{schannel}) into (\ref{W4final}) and get
\begin{equation}\label{W4schannel}
\begin{aligned}
W_4=&p^{-N\sum^4_{i=1}\Delta_i}\sum_{\ell}\frac{C_{12\ell}C_{\ell 34}}{|x_{12}|_p^{\Delta_1+\Delta_2-\Delta_{\ell}}|x_{13}|_p^{\Delta_{\ell}+\Delta_3-\Delta_{4}}|x_{14}|_p^{2\Delta_4}}
\end{aligned}
\end{equation}
To compare with the boundary four-point function, plug in the conformal block (\ref{CB}) in s-channel to the boundary  four-point function (\ref{4ptpadic}), we see that the Wilson line result (\ref{W4schannel}) reproduces exactly the 
boundary four-point function in the s-channel. 
The t-channel and u-channel are parallel.
From this computation we again see that the final result does not depend on the positions of the two internal vertices $v_a$ and $v_b$, due to the fact that the configuration (\ref{puregaugeEx}) is a pure gauge.
This is completely analogous to the $\mathbb{R}$ case.

\section{Connection with tensor network}

In Section~\ref{sec:WLN}.\ we have shown that the PGL$(2,\mathbb{Q}_p)$ Wilson lines with the connection (\ref{puregaugeEx}) (which is analogous to pure AdS solution) recovers correlation functions of a boundary $p$-adic CFT. 
Since a $(p+1)$-valent tree type tensor network can realize the $p$-adic AdS/CFT \cite{Bhattacharyya:2017aly}, one might ask what the relation between the Wilson line network and the tensor network (for the same $p$-adic AdS/CFT) is.
In this section we will show that the most natural interpretation of a tensor network that corresponds to a $p$-adic AdS/CFT is precisely as a Wilson line network. 

\subsection{Individual tensors and bulk WL junction with nearest neighbors}

In a tensor network realization of the $p$-adic AdS/CFT, the ``network" is the ($(p+1)$-valent) Bruhat-Tits tree and on each vertex of the tree sits a rank-$(p+1)$ tensor. 
All the tensors in the bulk are contracted with its nearest $(p+1)$ neighbors along the edges on the tree; whereas the tensors on the boundary of the tree\footnote{In practice the boundary is on a cut-off surface, thus is of finite distance away from the origin of the tree.} have uncontracted legs and correspond to the coefficients of boundary wavefunction or path integral, depending on whether the time direction is included or not \cite{Bhattacharyya:2017aly, us2}.

The central object is thus the individual tensor. 
As shown in  \cite{Bhattacharyya:2017aly, us2}, for a $p$-adic CFT with spectrum $\{\Delta_i\}$ and three-point coefficients $\{C_{ijk}\}$, the corresponding tensor network has tensor
\begin{equation}
T_3=p^{-\Delta_1-\Delta_2-\Delta_3}C_{123}
\end{equation}
Namely, each junction carries a structure coefficient $C_{ijk} $ with  each of its leg-$i$ carrying a factor $p^{-\Delta_i}$ (to measure the distance on the tree).
Now we show that this individual tensor can be matched to 
the smallest bulk Wilson line junction where the junction connects to its nearest neighbors.
 
In particular, recall that when contracting the Wilson line network onto representation $\langle \Delta |$, there is a free parameter $Z$ which originates from the overlap (\ref{overlap2}) and appears in expectation values of bulk Wilson line networks. 
In Section 4. since the goal is to reproduce the boundary $p$-adic correlation functions, the vertices are pushed all the way to the boundary and thus the $Z$ parameter drops out of the final result. 
Now, to match the bulk Wilson line junction (with vertices sitting in the bulk of the tree) with tensors in the interior of the tensor network, we find that $Z$ is no longer free, and there exists a natural choice of $Z$ such that the two networks match perfectly.

Let us revisit the evaluation of bulk Wilson line junction in Section~\ref{subsubsec:bulkWLJ}, in particular the computation in (\ref{3pt}). 
WLOG, we place the junction at the origin of the tree $\mathfrak{o}^{\textrm{BT}}$, with connection (\ref{OBT}). 
Now, consider  three Wilson lines $\hat{ \mathfrak{W}}_{\Delta_{1,2,3}}(\mathfrak{o}^{\textrm{BT}}\to v_{1,2,3})$ connecting $\mathfrak{o}^{\textrm{BT}}$ to three of its nearest neighbours $v_{1,2,3}$, with group element 
\be \label{neighg}
\mathfrak{g}^{\textrm{BT}}(v_{i}) \in \bigg\{ \Gamma_n \equiv \left(\begin{array}{cc} p& n\\ 0 & 1\end{array}\right)\bigg\vert_{n=0,\cdots p-1},\,\,\Gamma_{p} \equiv \left (\begin{array}{cc} 1& 0\\ 0 & p\end{array}\right)\bigg \}.
\ee
We find
\be
\left|\frac{\det \Gamma_{m<p}}{((\Gamma_{m<p})_{21}X + (\Gamma_{m<p})_{22})^2}\right|_p = p^{-1},  \qquad  \left|\frac{\det \Gamma_p}{((\Gamma_p)_{21}X + (\Gamma_p)_{22})^2}\right|_p = p. 
\ee
We denote this ``nearest neighbour junction'' by $T_3(m_1,m_2,m_3)$, where these $m_i$ labels the choice of neighbours with $g^{\textrm{BT}}$ taking values in (\ref{neighg}).  
Using also (\ref{3pt}), we find that if all three Wilson lines are ``climbing up'' the tree, i.e.\ $m<p$, the 3-pt junction gives \begin{equation}
\begin{aligned}
T_3(m_1,m_2,m_3) = &  \frac{ p^{-\Delta_1-\Delta_2-\Delta_3} C_{123}}{|m_1-m_2|_p^{\Delta_1+\Delta_2-\Delta_3} |m_2 - m_3|_p^{\Delta_2+\Delta_3-\Delta_1}|m_1-m_3|_p^{\Delta_1+\Delta_3-\Delta_2}}\\
&= p^{-\Delta_1-\Delta_2-\Delta_3} C_{123}, \qquad \textrm{for} \, n_{1,2,3} \neq p; \label{t31}
\end{aligned}
\end{equation}
And if one of the Wilson lines takes $m=p$ which ``climbs down'' the tree, we have
\begin{equation}
\begin{aligned}
&T_3(m_1,m_2,m_3=p)    \\
=& \frac{p^{-\Delta_1-\Delta_2-\Delta_3}C_{123}}{|m_1-m_2|_p^{\Delta_1+\Delta_2-\Delta_3} |p m_2 + (p^2 - 1)Z |_p^{\Delta_2+\Delta_3-\Delta_1}|p m_1 + (p^2 - 1)Z|_p^{\Delta_1+\Delta_3-\Delta_2}} \\
\end{aligned}
\end{equation}
which equals  (\ref{t31}) {\it  iff } 
\begin{equation}
|Z|_p=1
\end{equation}
With this choice, the nearest neighbour three-way Wilson line  junction has an expectation value that is in complete agreement with the prescription of the tensor network that recovers the correct correlation function of the $p$-adic CFT in \cite{Bhattacharyya:2017aly, us2}.

\subsection{Nearest neighbour $(p+1)$-point junction}
The computation above concerns only three-point junctions. Comparing with the tensor network where each tensor has $p+1$ legs, a three-point junction corresponds to projecting $(p-2)$ legs to the identity state with $\Delta=0$. 
Therefore we would like to generalize the above discussion to computing the nearest neighbour $(p+1)$-point junction.

Our intertwiners are defined explicitly for three-point junction in (\ref{intertwiner}). 
To deal with a generic $p+1$-point junction, our strategy is to mimic the computation of the four-point junction, and introduce auxiliary {\it virtual} Wilson lines that connect the $p+1$ point junction to itself. 
To decompose a $p+1$-point junction into a fusion tree of three-point junction, we will need $p-2$ virtual Wilson lines. 
As a result, a nearest neighbour $(p+1)$-point junction $T_{p+1}$ would take the following form
\begin{eqnarray}
&&T_{p+1} =  p^{-\Delta_1-\cdots \Delta_{p}  + \Delta_{p+1}} \sum_{\{I_a\}} \left( \prod_{a=1}^{p-2}  \int_{\mathbb{Q}_p} dy_a\right)   \langle \mathcal{O}_{1}(X_1) \mathcal{O}_{2}(X_2) \mathcal{\tilde O}_{{I_1}}(y_1) \rangle \nonumber \\
&& \times  \langle \mathcal{O}_{{I_1}}(y_1)\mathcal{O}_{3}(X_3)  \mathcal{\tilde O}_{{I_2}}(y_2) \rangle \cdots
 \langle \mathcal{O}_{{I_{p-2}}}(y_{p-2})\mathcal{O}_{p}(X_{p})  \mathcal{O}_{{p+1}}(X_{p+1}) \rangle,
\end{eqnarray}
where 
\be
X_i = \mathfrak{g}^{\textrm{BT}}(v_i) \cdot Z
\ee
Using (\ref{3ptO34}) and (\ref{3pt12shadow}), we have
\be
T_{p+1} = p^{-\sum_{i=1}^{p+1}\Delta_i} \sum_{I_{a_1}\cdots I_{a_{p-2}}} C_{12 {I_{a_1}}} C_{I_{a_1} 3I_{a_2}} \cdots C_{I_{a_{p-2}}p(p+1)}, \qquad \textrm{for}\,\, |Z|_p = 1.
\ee 
This result is in complete agreement with the full prescription of the tensor network for each tensor.
The method introduced here can be used for computing a generic $n$-point junction. 
This provides evidence that the tensor network can be interpreted as a Wilson line network. 
 
\section{Summary and Outlook}
\subsection{Summary}

We have used PGL$(2,\mathbb{Q}_p)$ Wilson line network to reproduce correlation functions of $p$-adic CFT.
For $p$-adic CFT with complex-valued fields, the PGL$(2,\mathbb{Q}_p)$ configuration is given by (\ref{puregaugeEx}), which is the analogue of the pure AdS solution of the $\mathfrak{sl}(2,\mathbb{C})$ Chern-Simons theory.

We also constructed a convenient set of basis for the representations of PGL$(2,\mathbb{Q}_p)$, inspired by the shadow operator formalism that is used extensively in the usual CFTs based on $\mathbb{R}$ or $\mathbb{C}$.
These representations are used to define Wilson lines and networks of Wilson lines. 
We have explicitly studied  networks with two, three, and four external lines, and shown that when the end-points of these external lines are pushed to the boundary of the tree, the expectation values of these Wilson line networks precisely reproduce the correlation functions of the dual $p$-adic CFT.  

Moreover, we have computed a network of $(p+1)$ Wilson lines where they meet at a common vertex and each with an open end located at a nearest neighbor of the meeting vertex. 
We have shown that (after a free parameter in the Wilson line formulation is fixed appropriately) it matches exactly with a prescription of the tensor network that recovers all the dual $p$-adic CFT correlation functions. 
This suggests an alternate interpretation of the tensor network in \cite{us2} as a network of Wilson lines.

\subsection{Outlook}
We have used the configuration (\ref{puregaugeEx}) that is the analogue of the pure AdS solution of the $\mathfrak{sl}(2,\mathbb{C})$ Chern-Simons theory. 
However, we have not completely defined the lattice gauge theory  which would have this configuration as a vacuum solution. 
One difficulty is that, in the usual formulation of a lattice Chern-Simons theory, the action fixes the flux threading every closed-loop to zero. 
Indeed, for a Chern-Simons theory to be defined on a lattice, the lattice needs an equal number of edges and loops.

In the current situation, our lattice gauge theory lives on a tree, i.e.\ without any loop, thus the Chern-Simons like theory that lives on the tree would need to be rather different from the usual lattice Chern-Simons theory.
We leave the full construction of this theory to future work.
One other possibility is to add links to form closed loops. 
The corresponding Chern Simons theory might then have more dynamics. 
It would be interesting to explore the possible relation between this theory and the bulk theory with edge dynamics considered in \cite{Gubser:2016htz}. 
Finally, it would be interesting to find the bulk dual of $p$-adic CFTs with $p$-adic valued fields (hence with descendants.)

%Another modification that should lead to interesting dynamics is to consider the version of $p$-adic CFT where descendants are allowed. 

One initial motivation to study the tensor network construction that inspired the current study is the problem of bulk reconstruction. 
In this paper, we have shown that a natural tensor network that recovers the correlation functions of the dual theory is more naturally interpreted as a network of Wilson lines. 
The success of the tensor network can be attributed to the strong constraints of symmetries, namely, operators/states that transform as representations of the conformal group are directly encoded in the bulk tensors, which naturally parallel the Wilson lines. 
In other words, the Wilson line network can be realized by a tensor network. 
Wilson line can be thought of as an example where the challenge of probing physics below the AdS scale in a tensor network (see e.g. \cite{Bao2015}) can be overcome by symmetries. 
This should lead to new insights in the tensor network bulk reconstruction program. 

Finally, it would also be interesting to explore to what extent the error correcting properties of tensor network (see e.g.\cite{Pastawski:2015qua, Hayden:2016cfa, Harlow}) can directly manifest themselves in the Wilson line network. 
We leave these interesting problems to future work. 

\section*{Acknowledgements} 
We thank Arpan Bhattacharyya and Long Cheng for initial collaboration and Matthias Gaberdiel for helpful discussions. 
LYH thanks ITP-CAS and WL thank Fudan University for hospitality during various stages of this project. 
We are grateful for support from the Thousand Young Talents Program.

\appendix

\section{A few identities for shadow operator}
\subsection{Orthogonality between operator and its shadow operator}
\label{append:OtO}
In this subsection we prove the orthogonality condition (\ref{OtO}) between an operator $\mathcal{O}_{\ell}$ and its shadow operator $\tilde{\mathcal{O}}_{\ell}$ (defined in (\ref{shadowdef})). 
The two-point function between $\mathcal{O}_{\ell}$ and its shadow $\tilde{\mathcal{O}}_{\ell}$ involves a $p$-adic integral
\begin{equation}\label{OtOintegral}
\langle \mathcal{O}_{\ell}(x_1) \tilde{\mathcal{O}}_{\ell}(x_2)\rangle =N(d,\Delta) \int_{\mathbb{Q}_p} dy \frac{|y-x_2|^{2\Delta_{\ell}-2d}}{|x_1-y|^{2\Delta_{\ell}}}
\end{equation}
where $N(d,\Delta)$ is the normalization constant, $d$ is the dimension and $d=1$ here.
The integral (\ref{OtOintegral}) can be evaluated using $p$-adic Fourier transform.

In $p$-adic Fourier transform, the analogue of $e^{2\pi i k x}$ 
is
\begin{equation}
\chi_{k}(x)\equiv e^{-2\pi i [k x]} \qquad \textrm{with}\quad k,x\in\mathbb{Q}_p
\end{equation}
where ``[x]" is the ``fractional part" of the $p$-adic number $x=\sum^{+\infty}_{n=-N}a_n p^n$, i.e,\ $[x]\equiv\sum^{-1}_{n=-N} a_n p^n$.
It is also called the additive character, and satisfies  \begin{equation}
\chi_{k}(x_1+x_2) = \chi_{k}(x_1) \chi_{k}(x_2)
\qquad 
\chi_{k_1+k_2}(x) = \chi_{k_1}(x) \chi_{k_2}(x)
\qquad 
\chi_{k}(0)=\chi_{0}(x)=1 
\end{equation}
The useful identity here is \cite{Koblitz}:
\begin{equation}\label{characteristic}
\int_{\mathbb{Z}_p} dx \chi_k(x)=\gamma(k)\equiv\begin{cases}
1 \qquad k\in\mathbb{Z}_p \\
0\qquad \textrm{otherwise}
\end{cases}
\end{equation}
where $\gamma(k)$ is the characteristic function of $\mathbb{Z}_p$.
Using (\ref{characteristic}), one can obtain the Fourier transform of $|x|^s_p$ 
\begin{equation} \label{Gubidentity}
\int_{\mathbb{Q}_p} \,dx \, \chi_{k}(x) |x|_p^{s} = \frac{\zeta_{p}(s+1)}{\zeta_p(s)}\frac{1}{|k|_p^{s+1}}+ \textrm{contact terms},
\end{equation}
where
$\zeta_p(s) \equiv \frac{1}{1-p^{-s}}$ is the $p$-adic Zeta function.\footnote{For more on $p$-adic Beta and Gamma functions, see e.g.\ \cite{GGP} and \cite{Koblitz}. } 
A double Fourier transform of (\ref{OtOintegral}) using (\ref{Gubidentity}) then gives
\be 
 \langle \mathcal{O}_{\ell}(x_1) \tilde{\mathcal{O}}_{\ell}(x_2)\rangle =  \delta(x_1-x_2) + \textrm{contact terms}, 
\ee
where we have used the normalization constant given in (\ref{normN}) and set $d=1$. 

\subsection{Three point function involving the shadow operator}
\label{append:3ptshadow}

In computing the three point function  (\ref{3pt12shadow}) between two operators $(\mathcal{O}_1$ and $\mathcal{O}_2$) and one shadow operator $\tilde{O}_{\ell}$, we first evaluated the OPE between $\mathcal{O}_1$ and $\mathcal{O}_2$ and then used the orthogonality condition (\ref{OtO}) between an operator and its shadow.
In this appendix, we would like to confirm this result by taking an alternative route. 

We will first plug in the definition of the shadow operator (\ref{shadowdef}) into the three-point function (\ref{3pt12shadow}) and then evaluate the three-point function of three ordinary operators:
\begin{equation}\label{3ptstep2}
\begin{aligned}
\langle \mathcal{O}_{1}(x_1) \mathcal{O}_{2}(x_2) \tilde{\mathcal{O}}_{\ell}(y) \rangle &=\int_{\mathbb{Q}_p}dz \, |z-y|_p^{2\Delta_\ell-2} \langle \mathcal{O}_{1}(x_1) \mathcal{O}_{2}(x_2) \mathcal{O}_{\ell}(z) \rangle\\
 &=\frac{C_{12\ell}}{|x_{12}|^{\Delta_1+\Delta_2-\Delta_{\ell}}} \int_{\mathbb{Q}_p}dz \, \frac{|z-(y-x_2)|_p^{2\Delta_\ell-2}}{|z|^{\Delta_2+\Delta_{\ell}-\Delta_1}_p \, |z-x_{12}|_p^{\Delta_\ell+\Delta_1-\Delta_2}}
 \end{aligned}
\end{equation}
where in the last step we have used the $p$-adic three-point function (\ref{3ptMobius}) fixed by $p$-adic M\"obius symmetry and finally shifted the integration parameter by $z\rightarrow z+x_2$. 
It then remains to evaluate the $p$-adic integral in (\ref{3ptstep2}).

Recall that the positions $x_1$ and $x_2$ are given, whereas $y$ still needs to be integrated over, therefore the natural scale in the integral in (\ref{3ptstep2}) is the norm $|x_{12}|_p$. 
Applying the ``isosceles property" on $|z-x_{12}|_p$, we can break the integral into two parts:
\begin{equation}\label{integralfull}
\begin{aligned}
&\mathcal{I}\equiv\int_{\mathbb{Q}_p}dz \, \frac{|z-(y-x_2)|_p^{2\Delta_\ell-2}}{|z|^{\Delta_2+\Delta_{\ell}-\Delta_1}_p \, |z-x_{12}|_p^{\Delta_\ell+\Delta_1-\Delta_2}}\\
&=\int_{|z|_p\geq |x_{12}|_p}dz \, \frac{|z-(y-x_2)|_p^{2\Delta_\ell-2}}{|z|^{2\Delta_{\ell}}_p}+\frac{1}{|x_{12}|_p^{\Delta_\ell+\Delta_{1}-\Delta_2}}\int_{|z|_p\leq |x_{12}|_p}dz \, \frac{|z-(y-x_2)|_p^{2\Delta_\ell-2}}{|z|^{\Delta_2+\Delta_{\ell}-\Delta_1}_p }
\end{aligned}
\end{equation}
where the first integral can be rewritten as
\begin{equation}
\int_{|z|_p\geq |x_{12}|_p}dz \, \frac{|z-(y-x_2)|_p^{2\Delta_\ell-2}}{|z|^{2\Delta_{\ell}}_p}=(\int_{\mathbb{Q}_p}dz-\int_{|z|_p\leq |x_{12}|_p}dz)  \, \frac{|z-(y-x_2)|_p^{2\Delta_\ell-2}}{|z|^{2\Delta_{\ell}}_p}
\end{equation}
Therefore the integral (\ref{integralfull}) contains two parts:
\begin{equation}
\mathcal{I}=\mathcal{I}_1+\mathcal{I}_2
\end{equation}
with
\begin{equation}\label{integral1}
\mathcal{I}_1\equiv \int_{\mathcal{Q}_p}dz \, \frac{|z-(y-x_2)|_p^{2\Delta_\ell-2}}{|z|^{2\Delta_{\ell}}_p}=\delta(y-x_2)+\textrm{contact terms}
\end{equation}
which, after combining with the other factor $\frac{C_{12\ell}}{|x_{12}|^{\Delta_1+\Delta_2-\Delta_{\ell}}}$, already gives us the result of three point function (\ref{3pt12shadow}).

Hence we need to explain the second part of the integral:
\begin{equation}\label{integral2}
\begin{aligned}
\mathcal{I}_2&=\frac{1}{|x_{12}|_p^{\Delta_\ell+\Delta_{1}-\Delta_2}}\int_{|z|_p\leq |x_{12}|_p}dz \, \frac{|z-(y-x_2)|_p^{2\Delta_\ell-2}}{|z|^{\Delta_2+\Delta_{\ell}-\Delta_1}_p }
-\int_{|z|_p\leq |x_{12}|_p}dz \, \frac{|z-(y-x_2)|_p^{2\Delta_\ell-2}}{|z|^{2\Delta_{\ell}}_p}
\end{aligned}
\end{equation}
which involves two integrals within the range $|z|_p \leq |x_{12}|_p$.
On the other hand, in the original three-point function (\ref{3pt12shadow}), the condition for evaluating OPE between $\mathcal{O}_1$ and $\mathcal{O}_2$ is that they are closer to each other than to $y$, hence we have
\begin{equation}
|z|_p \leq |x_{12}|_p \leq |y-x_2|_p
\end{equation}
which further simplifies the integral (\ref{integral2}) and gives
\begin{equation}
\mathcal{I}_2=\frac{|y-x_{2}|_p^{2\Delta_\ell-2}}{|x_{12}|_p^{2\Delta_\ell-1}} \left[ \zeta_p\left(1-(\Delta_2+\Delta_{\ell}-\Delta_1)\right)-\zeta_p\left(1-2\Delta_\ell\right) \right]
\end{equation}

Plug in the results of the two integrals back to  (\ref{3ptstep2}) we have
\begin{equation}
\langle \mathcal{O}_{1}(x_1) \mathcal{O}_{2}(x_2) \tilde{\mathcal{O}}_{\ell}(y) \rangle=\mathcal{K}_{1,2,\ell}(x_1,x_2,y)+\tilde{\mathcal{K}}_{1,2,\tilde{\ell}}(x_1,x_2,y)
\end{equation}
with
\begin{equation}\label{block}
\mathcal{K}_{1,2,\ell}(x_1,x_2,y)\equiv\frac{C_{12\ell}}{|x_{12}|_p^{\Delta_1+\Delta_2-\Delta_{\ell}}}\delta(y-x_2)
\end{equation}
extracting the contribution of $\mathcal{O}_{\ell}$ in the ($p$-adic) conformal block and
\begin{equation}
\tilde{\mathcal{K}}_{1,2,\tilde{\ell}}(x_1,x_2,y)\equiv\frac{C_{12\ell}|y-x_{2}|_p^{2\Delta_\ell-2}}{|x_{12}|_p^{\Delta_1+\Delta_2-(1-\Delta_{\ell})}} \left[ \zeta_p\left(1-(\Delta_2+\Delta_{\ell}-\Delta_1)\right)-\zeta_p\left(1-2\Delta_\ell\right) \right]
\end{equation}
is the extra term that arises from this route of computing three point function (\ref{3ptstep2}).
The interpretation of the $\tilde{\mathcal{K}}_{1,2,\tilde{\ell}}(x_1,x_2,y)$ is similar to its counterpart in the  $\mathbb{R}$ case \cite{SimmonsDuffin:2012uy}. 
Focus on the exponent of $|x_{12}|_p$, we see that $\tilde{\mathcal{K}}_{1,2,\tilde{\ell}}(x_1,x_2,y)$ is similar to $\mathcal{K}_{1,2,\ell}(x_1,x_2,y)$ except that the intermediate channel $\mathcal{O}_{\ell}$ is replaced by its shadow $\tilde{\mathcal{O}}_{\ell}$. 
This is precisely the analogue of the ``shadow block" that  appears in the case of $\mathbb{R}$ and should be discarded.

\bibliographystyle{utphys}
\bibliography{padic.bib}

\providecommand{\href}[2]{#2}\begingroup\raggedright\begin{thebibliography}{10}

\bibitem{Heydeman:2016ldy}
M.~Heydeman, M.~Marcolli, I.~Saberi, and B.~Stoica, ``{Tensor networks,
  $p$-adic fields, and algebraic curves: arithmetic and the AdS$_3$/CFT$_2$
  correspondence},'' \href{http://dx.doi.org/10.4310/ATMP.2018.v22.n1.a4}{{\em
  Adv. Theor. Math. Phys.} {\bfseries 22} (2018) 93--176},
\href{http://arxiv.org/abs/1605.07639}{{\ttfamily arXiv:1605.07639 [hep-th]}}.
%%CITATION = ARXIV:1605.07639;%%.

\bibitem{Gubser:2016guj}
S.~S. Gubser, J.~Knaute, S.~Parikh, A.~Samberg, and P.~Witaszczyk, ``{$p$-adic
  AdS/CFT},'' \href{http://dx.doi.org/10.1007/s00220-016-2813-6}{{\em Commun.
  Math. Phys.} {\bfseries 352} no.~3, (2017) 1019--1059},
\href{http://arxiv.org/abs/1605.01061}{{\ttfamily arXiv:1605.01061 [hep-th]}}.
%%CITATION = ARXIV:1605.01061;%%.

\bibitem{BruhatTits}
F.~Bruhat and J.~Tits, ``{Groupes r\'eductifs sur un corps local},'' {\em Inst.
  Hautes \'Etudes Sci. Publ. Math.} {\bfseries (41)} (1972) 5--251.

\bibitem{Zabrodin:1988ep}
A.~V. Zabrodin, ``{Nonarchimedean Strings and Bruhat-tits Trees},''
\href{http://dx.doi.org/10.1007/BF01238811}{{\em Commun. Math. Phys.}
  {\bfseries 123} (1989) 463}.
%%CITATION = CMPHA,123,463;%%.

\bibitem{Ostrowski}
A.~Ostrowski, ``{\"Uber einige L\"osungen der Funktionalgleichung $\phi(x)\cdot
  \phi(y)= \phi(xy)$},'' {\em Acta Mathematica (2nd ed.).} {\bfseries 41 (1)}
  271--284.

\bibitem{Gubser:2016htz}
S.~S. Gubser, M.~Heydeman, C.~Jepsen, M.~Marcolli, S.~Parikh, I.~Saberi,
  B.~Stoica, and B.~Trundy, ``{Edge length dynamics on graphs with applications
  to $p$-adic AdS/CFT},'' \href{http://dx.doi.org/10.1007/JHEP06(2017)157}{{\em
  JHEP} {\bfseries 06} (2017) 157},
\href{http://arxiv.org/abs/1612.09580}{{\ttfamily arXiv:1612.09580 [hep-th]}}.
%%CITATION = ARXIV:1612.09580;%%.

\bibitem{Gubser:2017vgc}
S.~S. Gubser, C.~Jepsen, S.~Parikh, and B.~Trundy, ``{O(N) and O(N) and
  O(N)},''
\href{http://arxiv.org/abs/1703.04202}{{\ttfamily arXiv:1703.04202 [hep-th]}}.
%%CITATION = ARXIV:1703.04202;%%.

\bibitem{Gubser:2017tsi}
S.~S. Gubser and S.~Parikh, ``{Geodesic bulk diagrams on the Bruhat-Tits
  tree},'' \href{http://dx.doi.org/10.1103/PhysRevD.96.066024}{{\em Phys. Rev.}
  {\bfseries D96} no.~6, (2017) 066024},
\href{http://arxiv.org/abs/1704.01149}{{\ttfamily arXiv:1704.01149 [hep-th]}}.
%%CITATION = ARXIV:1704.01149;%%.

\bibitem{Dutta:2017bja}
P.~Dutta, D.~Ghoshal, and A.~Lala, ``{Notes on exchange interactions in
  holographic p -adic CFT},''
  \href{http://dx.doi.org/10.1016/j.physletb.2017.08.042}{{\em Phys. Lett.}
  {\bfseries B773} (2017) 283--289},
\href{http://arxiv.org/abs/1705.05678}{{\ttfamily arXiv:1705.05678 [hep-th]}}.
%%CITATION = ARXIV:1705.05678;%%.

\bibitem{Gubser:2017qed}
S.~S. Gubser, M.~Heydeman, C.~Jepsen, S.~Parikh, I.~Saberi, B.~Stoica, and
  B.~Trundy, ``{Signs of the time: Melonic theories over diverse number
  systems},''
\href{http://arxiv.org/abs/1707.01087}{{\ttfamily arXiv:1707.01087 [hep-th]}}.
%%CITATION = ARXIV:1707.01087;%%.

\bibitem{Qu:2018ned}
F.~Qu and Y.-h. Gao, ``{Scalar fields on $p$AdS Scalar fields on $p$AdS},''
  \href{http://dx.doi.org/10.1016/j.physletb.2018.09.043}{{\em Phys. Lett.}
  {\bfseries B786} (2018) 165--170},
\href{http://arxiv.org/abs/1806.07035}{{\ttfamily arXiv:1806.07035 [hep-th]}}.
%%CITATION = ARXIV:1806.07035;%%.

\bibitem{Stoica:2018zmi}
B.~Stoica, ``{Building Archimedean Space},''
\href{http://arxiv.org/abs/1809.01165}{{\ttfamily arXiv:1809.01165 [hep-th]}}.
%%CITATION = ARXIV:1809.01165;%%.

\bibitem{Jepsen:2018dqp}
C.~B. Jepsen and S.~Parikh, ``{$p$-adic Mellin Amplitudes},''
\href{http://arxiv.org/abs/1808.08333}{{\ttfamily arXiv:1808.08333 [hep-th]}}.
%%CITATION = ARXIV:1808.08333;%%.

\bibitem{Gubser:2018cha}
S.~S. Gubser, C.~Jepsen, and B.~Trundy, ``{Spin in $p$-adic AdS/CFT},''
\href{http://arxiv.org/abs/1811.02538}{{\ttfamily arXiv:1811.02538 [hep-th]}}.
%%CITATION = ARXIV:1811.02538;%%.

\bibitem{Gubser:2018ath}
S.~S. Gubser, C.~Jepsen, Z.~Ji, and B.~Trundy, ``{Mixed field theory},''
\href{http://arxiv.org/abs/1811.12380}{{\ttfamily arXiv:1811.12380 [hep-th]}}.
%%CITATION = ARXIV:1811.12380;%%.

\bibitem{Orus:2013kga}
R.~Orus, ``{A Practical Introduction to Tensor Networks: Matrix Product States
  and Projected Entangled Pair States},''
  \href{http://dx.doi.org/10.1016/j.aop.2014.06.013}{{\em Annals Phys.}
  {\bfseries 349} (2014) 117--158},
\href{http://arxiv.org/abs/1306.2164}{{\ttfamily arXiv:1306.2164
  [cond-mat.str-el]}}.
%%CITATION = ARXIV:1306.2164;%%.

\bibitem{Pastawski:2015qua}
F.~Pastawski, B.~Yoshida, D.~Harlow, and J.~Preskill, ``{Holographic quantum
  error-correcting codes: Toy models for the bulk/boundary correspondence},''
  \href{http://dx.doi.org/10.1007/JHEP06(2015)149}{{\em JHEP} {\bfseries 06}
  (2015) 149},
\href{http://arxiv.org/abs/1503.06237}{{\ttfamily arXiv:1503.06237 [hep-th]}}.
%%CITATION = ARXIV:1503.06237;%%.

\bibitem{Hayden:2016cfa}
P.~Hayden, S.~Nezami, X.-L. Qi, N.~Thomas, M.~Walter, and Z.~Yang,
  ``{Holographic duality from random tensor networks},''
  \href{http://dx.doi.org/10.1007/JHEP11(2016)009}{{\em JHEP} {\bfseries 11}
  (2016) 009},
\href{http://arxiv.org/abs/1601.01694}{{\ttfamily arXiv:1601.01694 [hep-th]}}.
%%CITATION = ARXIV:1601.01694;%%.

\bibitem{Swingle:2009bg}
B.~Swingle, ``{Entanglement Renormalization and Holography},''
  \href{http://dx.doi.org/10.1103/PhysRevD.86.065007}{{\em Phys. Rev.}
  {\bfseries D86} (2012) 065007},
\href{http://arxiv.org/abs/0905.1317}{{\ttfamily arXiv:0905.1317
  [cond-mat.str-el]}}.
%%CITATION = ARXIV:0905.1317;%%.

\bibitem{Bhattacharyya:2017aly}
A.~Bhattacharyya, L.-Y. Hung, Y.~Lei, and W.~Li, ``{Tensor network and
  ($p$-adic) AdS/CFT},'' \href{http://dx.doi.org/10.1007/JHEP01(2018)139}{{\em
  JHEP} {\bfseries 01} (2018) 139},
\href{http://arxiv.org/abs/1703.05445}{{\ttfamily arXiv:1703.05445 [hep-th]}}.
%%CITATION = ARXIV:1703.05445;%%.

\bibitem{Hamilton:2005ju}
A.~Hamilton, D.~N. Kabat, G.~Lifschytz, and D.~A. Lowe, ``{Local bulk operators
  in AdS/CFT: A Boundary view of horizons and locality},''
  \href{http://dx.doi.org/10.1103/PhysRevD.73.086003}{{\em Phys. Rev.}
  {\bfseries D73} (2006) 086003},
\href{http://arxiv.org/abs/hep-th/0506118}{{\ttfamily arXiv:hep-th/0506118
  [hep-th]}}.
%%CITATION = HEP-TH/0506118;%%.

\bibitem{Hamilton:2006az}
A.~Hamilton, D.~N. Kabat, G.~Lifschytz, and D.~A. Lowe, ``{Holographic
  representation of local bulk operators},''
  \href{http://dx.doi.org/10.1103/PhysRevD.74.066009}{{\em Phys. Rev.}
  {\bfseries D74} (2006) 066009},
\href{http://arxiv.org/abs/hep-th/0606141}{{\ttfamily arXiv:hep-th/0606141
  [hep-th]}}.
%%CITATION = HEP-TH/0606141;%%.

\bibitem{Hamilton:2006fh}
A.~Hamilton, D.~N. Kabat, G.~Lifschytz, and D.~A. Lowe, ``{Local bulk operators
  in AdS/CFT: A Holographic description of the black hole interior},''
  \href{http://dx.doi.org/10.1103/PhysRevD.75.106001,
  10.1103/PhysRevD.75.129902}{{\em Phys. Rev.} {\bfseries D75} (2007) 106001},
  \href{http://arxiv.org/abs/hep-th/0612053}{{\ttfamily arXiv:hep-th/0612053
  [hep-th]}}.
[Erratum: Phys. Rev.D75,129902(2007)].
%%CITATION = HEP-TH/0612053;%%.

\bibitem{Gubser:1998bc}
S.~S. Gubser, I.~R. Klebanov, and A.~M. Polyakov, ``{Gauge theory correlators
  from noncritical string theory},''
  \href{http://dx.doi.org/10.1016/S0370-2693(98)00377-3}{{\em Phys. Lett.}
  {\bfseries B428} (1998) 105--114},
\href{http://arxiv.org/abs/hep-th/9802109}{{\ttfamily arXiv:hep-th/9802109
  [hep-th]}}.
%%CITATION = HEP-TH/9802109;%%.

\bibitem{Witten:1998qj}
E.~Witten, ``{Anti-de Sitter space and holography},''
  \href{http://dx.doi.org/10.4310/ATMP.1998.v2.n2.a2}{{\em Adv. Theor. Math.
  Phys.} {\bfseries 2} (1998) 253--291},
\href{http://arxiv.org/abs/hep-th/9802150}{{\ttfamily arXiv:hep-th/9802150
  [hep-th]}}.
%%CITATION = HEP-TH/9802150;%%.

\bibitem{us2}
L.-Y. Hung, W.~Li, and C.~M. Melby-Thompson, ``{$p$-adic CFT is a holographic
  tensor network},''
\href{http://arxiv.org/abs/1902.01411}{{\ttfamily arXiv:1902.01411 [hep-th]}}.
%%CITATION = ARXIV:1902.01411;%%.

\bibitem{Witten:1988hc}
E.~Witten, ``{(2+1)-Dimensional Gravity as an Exactly Soluble System},''
\href{http://dx.doi.org/10.1016/0550-3213(88)90143-5}{{\em Nucl. Phys.}
  {\bfseries B311} (1988) 46}.
%%CITATION = NUPHA,B311,46;%%.

\bibitem{Witten:2007kt}
E.~Witten, ``{Three-Dimensional Gravity Revisited},''
\href{http://arxiv.org/abs/0706.3359}{{\ttfamily arXiv:0706.3359 [hep-th]}}.
%%CITATION = ARXIV:0706.3359;%%.

\bibitem{Bhatta:2016hpz}
A.~Bhatta, P.~Raman, and N.~V. Suryanarayana, ``{Holographic Conformal Partial
  Waves as Gravitational Open Wilson Networks},''
  \href{http://dx.doi.org/10.1007/JHEP06(2016)119}{{\em JHEP} {\bfseries 06}
  (2016) 119},
\href{http://arxiv.org/abs/1602.02962}{{\ttfamily arXiv:1602.02962 [hep-th]}}.
%%CITATION = ARXIV:1602.02962;%%.

\bibitem{Besken:2016ooo}
M.~Besken, A.~Hegde, E.~Hijano, and P.~Kraus, ``{Holographic conformal blocks
  from interacting Wilson lines},''
  \href{http://dx.doi.org/10.1007/JHEP08(2016)099}{{\em JHEP} {\bfseries 08}
  (2016) 099},
\href{http://arxiv.org/abs/1603.07317}{{\ttfamily arXiv:1603.07317 [hep-th]}}.
%%CITATION = ARXIV:1603.07317;%%.

\bibitem{Fitzpatrick:2016mtp}
A.~L. Fitzpatrick, J.~Kaplan, D.~Li, and J.~Wang, ``{Exact Virasoro Blocks from
  Wilson Lines and Background-Independent Operators},''
  \href{http://dx.doi.org/10.1007/JHEP07(2017)092}{{\em JHEP} {\bfseries 07}
  (2017) 092},
\href{http://arxiv.org/abs/1612.06385}{{\ttfamily arXiv:1612.06385 [hep-th]}}.
%%CITATION = ARXIV:1612.06385;%%.

\bibitem{Ammon:2013hba}
M.~Ammon, A.~Castro, and N.~Iqbal, ``{Wilson Lines and Entanglement Entropy in
  Higher Spin Gravity},'' \href{http://dx.doi.org/10.1007/JHEP10(2013)110}{{\em
  JHEP} {\bfseries 10} (2013) 110},
\href{http://arxiv.org/abs/1306.4338}{{\ttfamily arXiv:1306.4338 [hep-th]}}.
%%CITATION = ARXIV:1306.4338;%%.

\bibitem{Castro:2018srf}
A.~Castro, N.~Iqbal, and E.~Llabr\'es, ``{Wilson lines and Ishibashi states in
  AdS$_{3}$/CFT$_{2}$},'' \href{http://dx.doi.org/10.1007/JHEP09(2018)066}{{\em
  JHEP} {\bfseries 09} (2018) 066},
\href{http://arxiv.org/abs/1805.05398}{{\ttfamily arXiv:1805.05398 [hep-th]}}.
%%CITATION = ARXIV:1805.05398;%%.

\bibitem{Besken:2018zro}
M.~Besken, E.~D'Hoker, A.~Hegde, and P.~Kraus, ``{Renormalization of
  gravitational Wilson lines},''
  \href{http://arxiv.org/abs/1810.00766}{{\ttfamily arXiv:1810.00766
  [hep-th]}}.

\bibitem{Kraus:2018zrn}
P.~Kraus, A.~Sivaramakrishnan, and R.~Snively, ``{Late time Wilson lines},''
\href{http://arxiv.org/abs/1810.01439}{{\ttfamily arXiv:1810.01439 [hep-th]}}.
%%CITATION = ARXIV:1810.01439;%%.

\bibitem{Koblitz}
{Koblitz, N}, {\em {$p$-adic Numbers, $p$-adic Analysis and Zeta-Functions}}.
\newblock {Springer}, 1994.

\bibitem{Gouvea}
{Gouv\^ea,~F.~Q.}, {\em {$p$-adic Numbers: An Introduction}}.
\newblock {Springer}, 1997.

\bibitem{Freund:1987kt}
P.~G.~O. Freund and M.~Olson, ``{NONARCHIMEDEAN STRINGS},''
\href{http://dx.doi.org/10.1016/0370-2693(87)91356-6}{{\em Phys. Lett.}
  {\bfseries B199} (1987) 186--190}.
%%CITATION = PHLTA,B199,186;%%.

\bibitem{Freund:1987ck}
P.~G.~O. Freund and E.~Witten, ``{ADELIC STRING AMPLITUDES},''
\href{http://dx.doi.org/10.1016/0370-2693(87)91357-8}{{\em Phys. Lett.}
  {\bfseries B199} (1987) 191}.
%%CITATION = PHLTA,B199,191;%%.

\bibitem{Brekke:1988dg}
L.~Brekke, P.~G.~O. Freund, M.~Olson, and E.~Witten, ``{Nonarchimedean String
  Dynamics},''
\href{http://dx.doi.org/10.1016/0550-3213(88)90207-6}{{\em Nucl. Phys.}
  {\bfseries B302} (1988) 365--402}.
%%CITATION = NUPHA,B302,365;%%.

\bibitem{Dragovich:2007wb}
B.~Dragovich, ``{Zeta strings},''
\href{http://arxiv.org/abs/hep-th/0703008}{{\ttfamily arXiv:hep-th/0703008
  [HEP-TH]}}.
%%CITATION = HEP-TH/0703008;%%.

\bibitem{Melzer:1988he}
E.~Melzer, ``{Nonarchimedean Conformal Field Theories},''
\href{http://dx.doi.org/10.1142/S0217751X89002065}{{\em Int. J. Mod. Phys.}
  {\bfseries A4} (1989) 4877}.
%%CITATION = IMPAE,A4,4877;%%.

\bibitem{SimmonsDuffin:2012uy}
D.~Simmons-Duffin, ``{Projectors, Shadows, and Conformal Blocks},''
  \href{http://dx.doi.org/10.1007/JHEP04(2014)146}{{\em JHEP} {\bfseries 04}
  (2014) 146},
\href{http://arxiv.org/abs/1204.3894}{{\ttfamily arXiv:1204.3894 [hep-th]}}.
%%CITATION = ARXIV:1204.3894;%%.

\bibitem{Banados:1998gg}
M.~Banados, ``{Three-dimensional quantum geometry and black holes},''
  \href{http://dx.doi.org/10.1063/1.59661}{{\em AIP Conf. Proc.} {\bfseries
  484} no.~1, (1999) 147--169},
\href{http://arxiv.org/abs/hep-th/9901148}{{\ttfamily arXiv:hep-th/9901148
  [hep-th]}}.
%%CITATION = HEP-TH/9901148;%%.

\bibitem{Li:2015osa}
W.~Li and S.~Theisen, ``{Some aspects of holographic W-gravity},''
  \href{http://dx.doi.org/10.1007/JHEP08(2015)035}{{\em JHEP} {\bfseries 08}
  (2015) 035},
\href{http://arxiv.org/abs/1504.07799}{{\ttfamily arXiv:1504.07799 [hep-th]}}.
%%CITATION = ARXIV:1504.07799;%%.

\bibitem{Bao2015}
N.~Bao, C.~Cao, S.~M. Carroll, A.~Chatwin-Davies, N.~Hunter-Jones, J.~Pollack,
  and G.~N. Remmen, ``{Consistency conditions for an AdS multiscale
  entanglement renormalization ansatz correspondence},''
  \href{http://dx.doi.org/10.1103/PhysRevD.91.125036}{{\em Phys. Rev.}
  {\bfseries D91} no.~12, (2015) 125036},
\href{http://arxiv.org/abs/1504.06632}{{\ttfamily arXiv:1504.06632 [hep-th]}}.
%%CITATION = ARXIV:1504.06632;%%.

\bibitem{Harlow}
D.~Harlow, ``{The Ryu-Takayanagi Formula from Quantum Error Correction},''
  \href{http://dx.doi.org/10.1007/s00220-017-2904-z}{{\em Commun. Math. Phys.}
  {\bfseries 354} no.~3, (2017) 865--912},
\href{http://arxiv.org/abs/1607.03901}{{\ttfamily arXiv:1607.03901 [hep-th]}}.
%%CITATION = ARXIV:1607.03901;%%.

\bibitem{GGP}
{Gelfand, I. M., Graev, M. I., Pyatetskii-Shapiro, I. I., Hirsch, K. A.
  (Translator)}, {\em {Representation Theory and Automorphic Functions}}.
\newblock {W. B. Saunders}, 1969.

\end{thebibliography}\endgroup

\end{document}